\documentclass[reprint,
prd,
aps,superscriptaddress,twocolumn,showpacs,nofootinbib,longbibliography]{revtex4-2}
\usepackage{graphicx,color}
\usepackage{amsmath,amsfonts,enumerate,amsthm,amssymb,bbm}
\usepackage{multirow}
\usepackage{bbold}
\usepackage{braket}
\usepackage{soul}
\usepackage{appendix}
\usepackage{scrextend}
\usepackage{xcolor}
\usepackage{caption}
\usepackage{subcaption}
\usepackage{graphicx}
\usepackage[colorlinks=true,citecolor=blue,linkcolor=magenta]{hyperref}

\mathchardef\ordinarycolon\mathcode`\:
\mathcode`\:=\string"8000
\begingroup \catcode`\:=\active
  \gdef:{\mathrel{\mathop\ordinarycolon}}
\endgroup

\theoremstyle{plain}

\theoremstyle{definition}

\theoremstyle{remark}

\def\<{\langle}

\def\H{ {\cal H} }

\def\>{\rangle}
\def\<{\langle}
\DeclareMathOperator{\Tr}{Tr}

\renewcommand{\ket}[1]{|#1\rangle}               

\newcommand{\edits}[1]{{\color[RGB]{0, 0, 0 }#1}}

\begin{document}

\title{A Search for Classical Subsystems in Quantum Worlds}

\author{Arsalan Adil}
\affiliation{Center for Quantum Mathematics \& Physics and Department of Physics \& Astronomy\\ UC Davis, One Shields Ave, Davis, CA.}
\affiliation{Basic Research Community for Physics e.V., Germany}

\author{Manuel S. Rudolph}
\affiliation{\'Ecole Polytechnique F\'ed\'erale de Lausanne, Lausanne, Switzerland}

\author{Andrew Arrasmith} 
\affiliation{
Theoretical Division, Los Alamos National Laboratory, Los Alamos, NM, USA.}

\author{Zoë Holmes}
\affiliation{\'Ecole Polytechnique F\'ed\'erale de Lausanne, Lausanne, Switzerland}

\author{Andreas Albrecht}
\affiliation{Center for Quantum Mathematics \& Physics and Department of Physics \& Astronomy\\ UC Davis, One Shields Ave, Davis, CA.}

\author{Andrew Sornborger}
\affiliation{
Information Sciences, Los Alamos National Laboratory, Los Alamos, NM, USA.}

\begin{abstract}
\noindent

Decoherence and einselection explain many features of the emergence classical world from an underlying quantum theory. However, the theory assumes a particular factorization of the global Hilbert space into constituent system and environment subsystems. In this work, given a fixed Hamiltonian we show that there can exist several factorizations (or tensor product structures) of a global Hilbert space that admit a quasi-classical description of subsystems in the sense that certain states (the ``pointer states'') are robust to entanglement. We derive several analytical forms that the Hamiltonian may take in such factorizations, each with its unique set of features. Our approach enables us to derive the division into quasi-classical subsystems and demonstrates that decohering subsystems do not necessarily align with our classical notion of locality. These results have interesting ramifications for the interpretation of quantum mechanics and, more practically, may be useful in 
algorithmically characterizing decoherence free subspaces.
\end{abstract}
\maketitle

\section{Introduction}
\label{sec:Intro}
To construct a  physical theory we usually start with a statement of which physical systems we intend to describe.  That choice, regardless of whether it is a harmonic oscillator, all known elementary particles, new hypothetical particles, or phonons in a condensed matter system, determines the mathematical scheme that describes the space of available states for the system.  With the state space in hand, we then build a full physical theory describing, for example, the system's evolution in time, its symmetries, etc. That is, we start with a notion of the relevant subsystems and then we use them to construct an appropriate Hamiltonian. 

However, is this too provincial an approach, born out of our familiarity with certain macroscopic objects? One could instead take the reverse approach and, starting from the global  Hamiltonian, ask if/what subsystems that system is composed of. This only entails choosing an energy spectrum and selecting associated eigenstates. Is it always possible, given an arbitrary Hamiltonian, to describe a system composed of various well-defined subsystems? Could more than one such composition coexist giving separate interpretations of the same global picture? To start with, this change in perspective poses the question, what do we actually mean by a subsystem? That is, given a potential factorization of the total Hilbert space, what properties should that factorization satisfy to define good subsystems?

Such questions have been explored by a number of authors using several different approaches~\cite{deutsch1985quantum-interpretation-basis,zanardi2001virtual-subsystems,
zanardi2004quantum-lloyd-lidar,
piazza2010glimmers, 
tegmark2015consciousness,
Cotler:2017abq,
carroll2021quantum-singh,
zanardi2022operational,
Stoica:2024nhh,zurek2025decoherence}. Here, we emphasize that these questions are related to the concepts of environment-induced superselection (einselection) and pointer states. \textit{Pointer states} are states of a subsystem that stay robust to entanglement with the rest of the physical world, namely, `the environment'. Often, but not always, these pointer states coincide with localized solutions to the Schr\"odinger equation and correspond to the types of entity that one usually associates with subsystems. \textit{Einselection} is the dynamical selection of these pointer states due to the environmental monitoring of the system \cite{zurek2003decoherence-and-the-quantum-origins, schlosshauer2019quantum-dec-review}. A paradigmatic example 
of this dynamical selection
would be the coherent states of a quantum harmonic oscillator interacting with a large environment \cite{paz1993reduction-zurek-habib, zurek1993coherent-states-Habib-Paz}. 

Pointer states have a special role in understanding decoherence due to several features (see e.g. Refs.~\cite{zurek1981pointer, joos1985emergence-zeh} for pioneering work and \cite{schlosshauer2019quantum-dec-review} for a review). \edits{One of these key features is that the pointer states evolve predictably (i.e. their evolution in the reduced system space is described by some suitable master equation) }~\cite{zurek1993preferred-predictability-sieve}. Furthermore, while the environment essentially performs a quantum non-demolition measurement on a system that is in a pointer state, systems in superpositions of pointer states rapidly decohere for systems in superpositions of pointer states, the environment acts as a ``which-path'' monitor and leads to the damping of interference terms \cite{schlosshauer2019quantum-dec-review}. 

However, despite its success in explaining several features of the quantum-to-classical transition, the decoherence program is built on the assumption that there is a particular ``splitting'' of the world into a system and an environment \textit{such that} there exist pointer states. 

As pointed out by Zurek \cite{zurek1998decoherence-existential-interpretation-rough-guide}, 
\begin{quote}
    …one issue which has often been taken for granted is looming large, as a foundation of the whole decoherence programme.  It is the question of what the `systems' which play such a crucial role in all the discussions of the emergent classicality are…  a compelling explanation of what the systems are -- how to define them given, say, the overall Hamiltonian in some suitably large Hilbert space -- would undoubtedly be most useful.
\end{quote}

It is this question that we set out to answer here. We take a deductive approach and ask, given a Hamiltonian, into what factors can the Hilbert space be split such that there exist pointer states? To understand this relation between pointer states and subsystems operationally, we introduce an algorithm which, given a Hamiltonian, $H$, as an input, seeks to find a system-environment split and its associated pointer states. We utilize a variational algorithm with a cost function that can be used to identify product states of a system which, when evolved under $H$, remain robust to entanglement. This operational perspective helps us identify subsystems that exhibit einselection. Furthermore, we discover several analytical forms of Hamiltonian that support einselection and categorize our numerical searches into these different forms.  

These analytical studies in conjunction with the numerical work using our algorithm have enabled us to identify different classes of phenomena (i.e. different forms of system-environment interaction and different families of pointer states) that satisfy our definitions. We have found that any Hamiltonian can admit a certain class of pointer states and system-environment splits built in a straightforward way from the eigenstates of the Hamiltonian.  This class is interesting in its own right, and serves as a useful reference point as we explore other ways our requirements can be realized.  We systematically delineate these classes in Section~\ref{sec:SR}. Some interesting special cases include 
solutions  where a pointer state exists for some, but not all, environment states; or that a certain state stays robust to entanglement longer than the typical decoherence time but its corresponding orthogonal state does not share this behavior.  

We have found that one can identify different system-environment splits that coexist within a given global system.  This observation may be relevant for engineering quantum technologies (for example in identifying all the available decoherence-free subspaces of specific quantum systems). We also comment on interesting foundational implications that arise in the cosmological context where there is no external ``observer''.

It is important to note that decoherence and einselection do not embody all the elements of the emergence of classical from quantum. Some of these points are made for example in Refs.~\cite{albrecht1994inflation-squeezing,Anglin:1996bb,polarski1996semiclassicality-starobinsky-decoherence-wo-decoherence,strasberg2023everything}. Furthermore, several aspects of the emergence of classical from quantum depend on a multipartite split to become fully realized.  For example, the Quantum Darwinism program~\cite{zurek2009quantum-darwinism} recognizes that the environment actively ``acts as a witness'' by redundantly storing information about the state of a system. Since individual observers only probe limited fragments of the environment, this selective redundancy of the pointer states is what leads to the emergence of  an ``objective'' classical reality.\footnote{Here ``objective'' reality is defined as the notion that multiple initially oblivious observers agree on the state of the system without affecting the others' outcomes ~\cite{zurek2009quantum-darwinism, ollivier2004objective, ollivier2005environment}.}
Still, decoherence phenomena as realized in bipartite system-environment factorization is an essential ingredient for most of such discussions, and that is the focus of our work here. In the conclusions we discuss how our work could be extended to the multipartite case.

\edits{We also note that many of the topics we discuss here can be explored using the consistent histories formalism~\cite{Griffiths:1984rx,
Omnes:1988ek,
Omnes:1988em,
Omnes:1988ej,
Omnes:1988fv,Omnes:1992ag,Gell-Mann:1992wkv,Dowker:1994ac,Gell-Mann:2011rif,arrasmith2019variational}.   At the end of Sect.~\ref{sec:relative-state} we connect specific results from this paper with earlier work using consistent histories. }


This paper is structured as follows. In Section~\ref{sec:what-is-a-subsytem}, we introduce the 
analytical 
notions of subsystems and pointer states. In Section~\ref{sec:methodology}, we introduce our algorithm for finding pointer states. In Section~\ref{sec:SR}, we provide a detailed analysis of our numerical and analytical findings. Section~\ref{sec:relative-state} explores the implications of our results for certain interpretational questions.  In Section~\ref{sec:conclusions} we digest our results and discuss what conclusions can be drawn. \edits{Appendix~\ref{sec:appendix-Definitions} clarifies our usage of certain terminology.} Appendix~\ref{sec:appendix-BlockD} gives the details of a specific block-diagonal form for the Hamiltonian which we show is possible for any Hamiltonian given suitably chosen subsystems. Appendix~\ref{sec:numerical-implementation} contains details regarding the implementation of the numerical algorithm. Appendix~\ref{sec:appendix-violin-plots} gives details of the statistics that emerge from our numerical search algorithm. 
This work has a broad range of motivations, from understanding the emergence of fundamental laws of physics, to discovering technical results which may be relevant for quantum technologies.  We gather these reflections in Section~\ref{sec:conclusions}. 

\section{What is a subsystem?} 
\label{sec:what-is-a-subsytem}

Typically, in order to frame a problem of interest, we assume a particular factorization in the underlying Hilbert space. Conventionally, one considers building a space as a tensor product of Hilbert space factors $\mathcal{H_A}$ and $\mathcal{H_B}$. 
One can thus form a basis $\{\ket{k}_w\}$ for ``world'' $w$ out of bases $\{\ket{i}_{\mathcal{A}}\}$ and $\{\ket{j}_{\mathcal{B}}\}$ (spanning $\mathcal{H_A}$ and $\mathcal{H_B}$ respectively) as
\begin{equation}
    \ket{k}_{w} \equiv \ket{i(k),j(k)}_{w} \equiv \ket{i}_{\mathcal{A}} \ket{j}_{\mathcal{B}}
    \label{eq:w2ab}
\end{equation}
defined by the mappings $i(k)$ and $j(k)$.
\footnote{Here, $i(.)$ and $j(.)$ should be thought of as surjective functions that map the indices $k\in \{1,...,d_w=d_\mathcal{A} d_\mathcal{B}\}$ to $i(.) \in \{1,...,d_\mathcal{A}\}$ and $j(.) \in \{ 1,...,d_\mathcal{B}\}$ where $d_X$ denotes the dimension of the relevant Hilbert space. Conventionally, in quantum information theory, one uses the mapping $k = i d_\mathcal{B} + j$, treating instead $i,j$ as free indices, but that is only one possible choice for the mapping.}

For this work we consider the reverse process - we start with some basis $\{\ket{k}_{w}\}$ that spans ${w}$ and use Eq.~\eqref{eq:w2ab} to define the tensor product structure (TPS). Operating in this way, we could alternatively insert
\begin{equation}
    \ket{k}_{w}' \equiv B\ket{k}_{w} 
\end{equation}
in the left side of Eq.~\eqref{eq:w2ab}, where $B$ is some unitary on ${w}$.  This scheme generates a different TPS (or factorization) determined by $B$.  In what follows we will use the set of all unitaries on ${w}$ to scan through different TPS's.  We consider only finite systems and limit our explorations to bipartite factorizations. 

Operationally, the TPS arises in the various operators one constructs to do physics, such as observables and parts of the Hamiltonian that act on, say, subsystem $\mathcal{A}$ or subsystem $\mathcal{B}$, exclusively. Note that operators that are ``local'' in the $\mathcal{H_A}\otimes\mathcal{H_B}$ factorization need not stay local from the point of view of a different factorization $\mathcal{H_{A'}}\otimes\mathcal{H_{B'}}$. 
 
Formally, using $\mathcal{H}$ to denote a Hilbert space, we write 
\begin{equation}
    \mathcal{H}_w \equiv \mathcal{H_A} \otimes \mathcal{H_B}.
\end{equation}
This particular factorization can be related to another one by a unitary transformation $B$,
\begin{equation}
    \label{eq:B-def}
    \mathcal{H_{A}} \otimes \mathcal{H_{B}} \xrightarrow{B} \mathcal{H_{A'}} \otimes \mathcal{H_{B'}}
\end{equation}
When applied to the Hamiltonian, the transformation $BHB^\dagger$ leaves the eigenvalues invariant, though the eigenstates of $H$ can look very different when expressed in different bases reflecting different factorizations.

At the level of the state, the change of TPS reflects a change in the entanglement structure between the subsystems described by a ket, $|\psi\rangle_w$. Since entanglement is factorization dependent, a product state, will generally be entangled in another arbitrary factorization. That is,
\begin{equation}
    \label{eq:psiw-lab}
    |\psi\rangle_w = \sum_{i,j} a_i |i\rangle_\sigma \otimes b_j |j\rangle_\epsilon = \sum_{i,j} \alpha_{ij} |i\rangle_s \otimes |j\rangle_e
\end{equation}
where, for the rest of this paper and in keeping with the decoherence literature, we use the notation $\sigma/s$ and $\epsilon/e$ to label the ``system'' and ``environment'' Hilbert spaces, respectively. In this notation, Eq.~\eqref{eq:B-def} becomes $\mathcal{H}_\sigma \otimes \mathcal{H}_\epsilon \xrightarrow{B} \mathcal{H}_s \otimes \mathcal{H}_e$ . The basis $|i\rangle_s$ is, in general, not related by a unitary transformation to $|k\rangle_\sigma$ (and likewise for the environment bases) even though clearly the global bases in each representation of $\H_w$ are related by the global unitary, $B$.\footnote{Intuitively, this is much like the case in $\mathbb{R}^3$, where a coordinate system $\{\hat{e}^I \}$ can be rotated by a transformation $M\in SO(3)$ to a basis $\{\hat{\theta}^I \}$ though individual basis elements in each coordinate basis will not generally be related by a lower dimensional isometry in $SO(2)$.} 

As a concrete example of this phenomenon one can consider the Bell states which are entangled in the $s$ and $e$ factorization but, for example, could be non-entangled in the $\sigma$ and $\epsilon$ factorization:
\begin{equation}
\begin{aligned}
     & |\Phi_+\rangle_w = \frac{1}{\sqrt{2}} \left( |0 \rangle_s \otimes   |0 \rangle_s   + |1 \rangle_s \otimes   |1 \rangle_s \right) \equiv |0\rangle_\sigma \otimes |0\rangle_\epsilon \\ 
     & |\Phi_-\rangle_w = \frac{1}{\sqrt{2}} \left( |0 \rangle_s \otimes   |1 \rangle_s   - |1 \rangle_s \otimes   |1 \rangle_s \right) \equiv |0\rangle_\sigma \otimes |1\rangle_\epsilon  \\
     & |\Psi_+\rangle_w = \frac{1}{\sqrt{2}} \left( |0 \rangle_s \otimes   |1 \rangle_s   + |1 \rangle_s \otimes   |0 \rangle_s \right) \equiv |1\rangle_\sigma \otimes |0\rangle_\epsilon  \\
     & |\Psi_-\rangle_w = \frac{1}{\sqrt{2}} \left( |0 \rangle_s \otimes   |1 \rangle_s   - |1 \rangle_s \otimes   |0 \rangle_s \right) \equiv |1\rangle_\sigma \otimes |1\rangle_\epsilon  \, .
\end{aligned}
\end{equation}
That is, here we have chosen the $\sigma$ and $\epsilon$ tensor product structure such that the Bell states are product states. Correspondingly, the unitary $B$ that transforms from the computational basis of $s$ and $e$, i.e., $\{ |0\rangle_s\otimes 0 \rangle_e , |0\rangle_s \otimes 1 \rangle_e, |1\rangle_s \otimes 0 \rangle_e, |1\rangle_s \otimes 1 \rangle_e \} $, to the Bell basis, $\{ |\Phi_+\rangle_{se} = |0\rangle_\sigma \otimes |0\rangle_\epsilon, |\Phi_-\rangle_{se} = |0\rangle_\sigma \otimes |1\rangle_\epsilon, |\Psi_+\rangle_{se} = |1\rangle_\sigma \otimes |0\rangle_\epsilon, |\Psi_-\rangle_{se} = |1\rangle_\sigma \otimes |1\rangle_\epsilon \}$ relates these two factorizations. However, different choices in $B$ would lead to different factorizations.

This begs the question: why has the physicist chosen a particular factorization? Is it merely out of convenience? Or to represent obliviousness (perhaps lack of control) towards one of the subsystems? Or perhaps that, as classical macroscopic subsystems operating in a warm environment, we are accustomed to identifying other classical subsystems at the macroscopic scale and that we are exporting this classical intuition to the quantum scale? 

Here, we explore a notion of subsystems associated with the processes of decoherence and einselection. 
Concretely, we ask: starting from only the total energy eigenvalues, can we arrive at a ``preferred'' factorization where the dynamics of the world (i.e. the system plus the environment) admit states in the system Hilbert space that stay robust to entanglement with the environment?

Some have argued that these preferred bipartite factorizations can be identified by their propensity to admit localized solutions \cite{piazza2010glimmers}.  While this certainly coincides with what we experience classically, this criterion is insufficient for identifying preferred factorizations though it is possible that these ``localized''\footnote{This strictly means that $\Delta x$, the width of the wavefunction, remains small} solutions emerge as by-products of other underlying physical mechanisms in certain settings. 

Consider, for example, the Adapted Caldeira-Leggett (ACL) model of Ref.~\cite{albrecht2023adaptedcalmodel} where a quantum simple harmonic oscillator (SHO) interacts with a random environment so that the Hamiltonian takes the form, 
\begin{equation}
    H_w = \alpha_s H_{\rm SHO}\otimes 1_e + \beta q_{\rm SHO} \otimes H_{e}^{\rm int} + \alpha_e 1_s \otimes H_e^{\rm self} \,.
\end{equation}
Then, in the strong interaction limit ($\alpha_i \ll \beta$), the eigenstates of the position operator serve as states that remain robust to entanglement with the environment; in the weak interaction limit ($\beta \ll \alpha_i$), where the self-Hamiltonian dominates, the energy eigenstates of the SHO remain robust to entanglement; finally, in the intermediate case, the coherent states (eigenstates of the annihilation operator) emerge as the pointer states \cite{zurek1993coherent-states-Habib-Paz}. Of these, the second case, barring perhaps the ground state wavefunction, is explicitly non-local even though the Hamiltonian takes on a form that coincides with the typical description of a weakly coupled system. Thus, although pointer states {\it exist} in all three regimes of coupling strengths, they are ``local'' only in the strong and intermediate regimes (see~\cite{Paz_1999} for a related discussion). The admission of local states must therefore be a consequence, not the underlying cause, of the emergence of preferred factorizations of the Hilbert space. 

\section{An algorithmic search for subsystems}
\label{sec:methodology}
We now make precise how one can iterate over different factorizations of the Hilbert space to algorithmically discover the quasi-classical subsystems of a Hamiltonian. 
\subsection{The optimization problem}
This search is, in principle, straightforward: wander in the space of unitaries, $B$, each of which defines a new factorization for the Hilbert space; for each choice of $B$ iterate over the space of initial states, $|\psi\rangle_w$; and for each choice of $|\psi\rangle_w$ evolve your system to late times and check if your reduced density matrix stably exhibits high values of purity. 
Thus, to identify whether an initial state, $|\psi\rangle_w$,
evolving under a Hamiltonian, $H_w$, admits a well-defined subsystem, the question is whether we can identify some basis, mathematically denoted by the  unitary rotation $B$ from the working basis, such that on rotating the composite system, $\rho_w(t)$, by $B$ \textit{for all times $t \leq T$}, the reduced state is pure~\footnote{We emphasize that the pointer states need not form a complete basis (as illustrated for example by the ``furnace'' case in subsection~\ref{subsec:Furnace}).  The complete basis denoted by $B$ lives in the global Hilbert space and is used to define the system-environment split.}. 
More concretely, we consider the linear entanglement entropy of the subsystem 
\begin{equation}
    \label{eq:lin-entropy}
    S(|\psi\rangle_w ) :=  1-\Tr_s\left(\Tr_e[ \rho_w(0)]^2\right),
\end{equation}
of an initially pure world state,
\begin{equation}\label{eq:rhowpure}
    \rho_w(0) :=  |\psi\rangle \langle\psi|_w \, ,
\end{equation}
in the basis determined by $B$,
\begin{equation}\label{eq:rhoS}
    \rho_w^B(t) := B e^{-iHt}\rho_w(0)e^{iHt} B^\dagger \, .
\end{equation}
If the time averaged entropy of the world in basis $B$ over some time period $[0,T]$ (for some $T$ longer than the decoherence time) is small, i.e., if 
\begin{equation}\label{eq:ContCostB}
   \left\langle S \left(\rho_w^B(t)\right) \right\rangle := \frac{1}{T} \int_{0}^{T} S\left(\rho_w^B(t)\right) \, dt \, \ll  \, 1 \, ,
\end{equation}
we say that it is \textit{robust to entanglement}. The initial state of the system, i.e. $\Tr_e[\rho_w^B(0)]$, which is robust to entanglement for a $B$, $H$, and $|\psi\rangle_w$ such that Eq.~\eqref{eq:ContCostB} holds, is identified as the \textit{pointer state}. Any system that admits pointer states we will say is a \textit{subsystem}\footnote{We stress that our use of the term `subsystem' includes the notion that it is a \textit{good} subsystem - namely one that for certain initial states is robust to entanglement.}.

This, in principle, is the optimization problem to be solved. 
For practical reasons we use this discrete version of Eq.~\eqref{eq:ContCostB}:
\begin{equation}\label{eq:DisCostB}
   C(|\psi(0)\rangle_w, B) := \frac{\Delta t}{T_{\rm train}} \sum_{k=0}^{T_{\rm train}/\Delta t} \left( 1 - \Tr[\rho_s^B(k \Delta t)^2] \right) \, ,
\end{equation}
where $\Delta t$ is a finite time interval and $T_{\rm train}$ is a maximum training time.
Eq.~\eqref{eq:DisCostB} serves as the cost function that is to be minimized, by optimizing the initial state $|\psi(0)\rangle_w$ and the factorization $B$ for a given $H$, in our hunt for subsystems.
We delineate different approaches to the initial conditions in the following subsection.

It may not always be possible to achieve a vanishing cost (in fact, numerically this will essentially never occur), in which case we adopt the ``predictability sieve'' criterion~\cite{zurek1993preferred-predictability-sieve, zurek1993coherent-states-Habib-Paz} that is a weaker definition of subsystems: factorizations that admit \textit{approximate} pointer states in the system space that are weakly, though not entirely, robust to entanglement with the environment.

We implement techniques from density matrix renormalization group theory to optimize the initial state and factorization (both parameterized by unitaries) the details of which can be found in Appendix~\ref{subsec:appendix-numerical-imp-dmrg}. However, more broadly, our general algorithmic approach, namely minimizing Eq.~\eqref{eq:DisCostB}, is flexible and could be implemented in a number of different ways. Indeed, with the development of quantum hardware, the cost could eventually be measured directly on a quantum computer. Nonetheless, a reader interested in the implementation details of our optimization algorithm may refer to Appendix~\ref{subsec:appendix-numerical-imp-dmrg}.

In our investigations, we restrict ourselves to only bi-partite factorizations of finite-dimensional Hilbert spaces that are constructed from two-state systems (qubits) so that $d_w \equiv \dim(\mathcal{H}_w) = 2^{n_w}$, where $n_w$ is the total number of qubits. We also restrict our exploration to fixed sizes of the individual system and environment Hilbert space with dimensions $2^{n_s}$ and $2^{n_e}$ respectively. 

\subsection{Approaches to initial conditions}
\label{subsec:SR-initial-conds}
While we are interested in finding the optimal factorization, parameterized by the unitary $B$, our cost, Eq.~\eqref{eq:DisCostB}, is a function of both $B$ and the initial state $|\psi(0)\rangle_w$. Thus, while we always optimize over $B$, we explore various levels of optimizing the initial state which may correspond to different physical situations. We outline each approach below in increasing order of complexity. As noted in Section~\ref{sec:Intro}, this work has a range of motivations from constructing decoherence-free subspaces for quantum technologies to understanding the quantum-to-classical transition on cosmological scales. Thus, we also elaborate on how these different motivations are materialized in our choices for the initial conditions below. \\ 

\paragraph*{\textbf{No initial state optimization.}} Here we fix our initial state to be a product state in the fiducial basis $|\psi(0)\rangle_w = |\xi\rangle_\sigma  |\phi\rangle_\epsilon$ and optimize over just the factorization parameterized by $B$.

\paragraph*{\textbf{ System state optimization.}} In addition to optimizing over $B$, we also optimize over the initial state in the {\it fiducial} system Hilbert space, $\mathcal{H}_\sigma$, i.e. $|\psi(0)\rangle_w = A_\sigma \otimes 1_\epsilon (|\xi\rangle_\sigma  |\phi\rangle_\epsilon) $ where $A_\sigma \in U(d_s)$ is to be optimized.

\paragraph*{\textbf{ Global state optimization.}} In this approach, in addition to optimizing over the factorization parameterized by $B$, we optimize the global initial state, i.e $|\psi(0)\rangle_w = A_w |\xi\rangle_\sigma |\phi\rangle_\epsilon $ where $A_w \in U(d_w)$. \\

The first of the three initialization schemes is to address whether a set of dynamics, with a \textit{particular} initial condition, can be viewed as corresponding to two approximately well-defined subsystems. 

The second case, where only the system state in the \textit{fiducial} TPS can be varied, corresponds to the case encountered in the lab: the experimentalist can set the state of the system being studied, for example the spin of an electron, and not of the more disordered environment with which it interacts.

Our final optimization scheme is aimed at relieving ourselves of any prior notion of a working basis and, in cases where a notion of a system-environment factorization is not defined \textit{a priori}, allows us to \textit{define} the factorization in which a random initial system state decoheres. This last case especially pertains to cosmology\footnote{As an illustration, there are two ambiguities that arise in studying the decoherence of curvature perturbations that are setup during the inflationary epoch. One is in deciding the cut-off scale for Fourier modes that form the system versus the environment. For example, some have defined the modes that leave the Hubble horizon (i.e. the ones that leave observable imprints on large-scale structure) as the ``system'' while modes that stay within the Hubble horizon as the ``environment'', though this choice remains arbitrary \cite{kiefer1998emergence, burgess2006decoherence-holman,martineau2007decoherence,sharman2007decoherence,
burgess2023minimal-holman} (see also \cite{polarski1996semiclassicality-starobinsky-decoherence-wo-decoherence} where it is argued that it is not necessary to specify an ``environment'' in order to explain the emergent classicality for inflationary perturbations which is due to quantum squeezing \cite{albrecht1994inflation-squeezing} in what the authors refer to as \textit{``decoherence without decoherence''}. The second ambiguity is in deciding the initial state in which the fluctuations originate. This is often taken to be the canonical ``Bunch-Davies'' vacuum~\cite{bunch1978quantum}, but there are other choices (which, in some sense, may even be more ``natural'' and may have differing observable effects~\cite{Adil2023entanglement}).}, where there is no reason that a fiducial factorization should be privileged over another one.

\section{Results}
\label{sec:SR}
We now turn to our main findings from the computational and analytic explorations. We focus on four types of Hamiltonian and, for the numerical work, use all three initialization approaches delineated in Sec.~\ref{subsec:SR-initial-conds} for each Hamiltonian. Sec.~\ref{subsec:SR-results} summarizes results from the algorithm (with detailed statistics deferred to Appendix~\ref{sec:appendix-violin-plots}) and Sec.~\ref{sec:analysis} identifies these numerical results with analytic forms that the Hamiltonian takes in the new TPS.

In what follows, it is important to remember that there are two factorizations of the Hilbert space that one must keep track of: the ``fiducial'' factorization, $\mathcal{H_\sigma}\otimes \mathcal{H_\epsilon}$ in which the Hamiltonian is first supplied, and the ``destination'' $\mathcal{H}_s\otimes\mathcal{H}_e$ which is the one we land on. The two are related by the unitary $B$ as in Eq.~\eqref{eq:B-def}.

\subsection{Hamiltonians considered}
\label{subsec:SR-Hamiltonians-considered}
The existence of a subsytem description depends on the dynamics governing the evolution of the global state. The discussion of the sort of dynamics that enable the existence of pointer states is deferred to  Section~\ref{sec:analysis}; here we merely state our trial Hamiltonians. Note that a particular Hamiltonian is fundamentally characterized by its eigenvalues and any physical description is attributed to the choice of a fiducial factorization. \\

\paragraph*{\textbf{ Central Spin.}} Here we study a system with a central qubit in $\mathcal{H_\sigma}$ that interacts with a bath of environment qubits in $\mathcal{H_\epsilon}$, where the interaction with each environment qubit consists of a single Pauli string, in the presence of an external field. That is,
\begin{equation} \label{eq:central-spin}
H = \alpha \sigma^{(Z)}_s + \alpha \sum_i^{n_e} \sigma_{e_i}^{(Z)} + \sum_i^{n_e} \beta_i \sigma_s^{(p_i)} \otimes \sigma_{e_i}^{(q_i)}
\end{equation}
where $\sigma^{(.)} \in \{\sigma^X,\sigma^Y,\sigma^Z\}$ is a single qubit Pauli operator and we have suppressed the identity operators in each term. The central spin model is of interest to us because its parameters can be easily modified to study various cases that appear in the decoherence literature. For example, setting $\beta_i = 0$ gives a perfectly decoupled Hamiltonian or, by choosing each $p_i$ from $\{X,Y,Z\}$ such that $p_i \neq p_j$ (for $n_e \leq 3$) one can have a simple Hamiltonian which nonetheless contains several non-commuting terms. We studied both of these sub-categories with our algorithm. \\

\paragraph*{\textbf{ Quantum Measurement Limit.}} This Hamiltonian takes the form $H= H_\sigma \otimes H_\epsilon$ in the fiducial factorization where $H_{\sigma/\epsilon}$ are random (traceless) Hermitian matrices. \footnote{We use the term ``quantum measurement limit'' here, although the tensor product form is a special case of this limit (which is discussed in more generality in~\cite{schlosshauer2007decoherence}).}\\

\paragraph*{\textbf{ Random Hamiltonian.}} In this case $H_w$ is a randomly generated Hermitian matrix.

\subsection{Numerical results for the twelve combinations of Hamiltonians \& initial conditions}
\label{subsec:SR-results}
\begin{figure}
    \centering
    \includegraphics[width=0.5\textwidth]{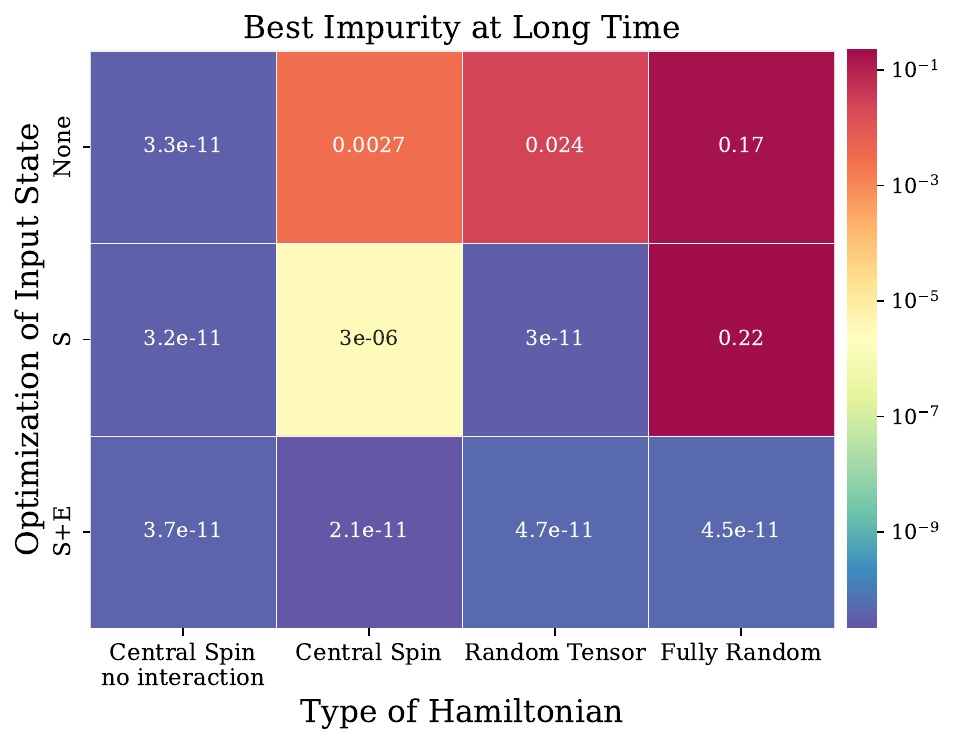}
    \caption{Best late time linear entropy achieved by the different optimization methods for various Hamiltonians. The columns correspond to the Hamiltonians described in Sec. \ref{subsec:SR-Hamiltonians-considered}. The rows correspond to the different \textit{initial state} optimizations from Sec.~\ref{subsec:SR-initial-conds}; the unitary $B$ is optimized in each case.  }
    \label{fig:algorithm_summary}
\end{figure}
We ran the four Hamiltonians in Sec.~\ref{subsec:SR-Hamiltonians-considered}, for each of the three approaches to initial conditions in Sec.~\ref{subsec:SR-initial-conds}, $\mathcal{O}(100)$ times to explore the solution space obtained by our algorithm. For each iteration, we generate a starting guess randomly for the unitaries $B$ (parameterizing the TPS) and $A$ (parameterizing the initial state; as needed), as well as generating a new Hamiltonian for the random Hamiltonian and quantum measurement limit cases. We summarize our results in the tile plot of Fig.~\ref{fig:algorithm_summary} where we select the run achieving the lowest cost in the late time limit for each of the categories. We take this late-time limit to be $t\approx 10^{5.5}$ due to numerical considerations (as discussed in Appendix~\ref{sec:appendix-violin-plots}). Let us understand these statistics row-by-row. 

The top row of Fig.~\ref{fig:algorithm_summary} shows results for the case where there is no optimization over the unitary $A$ (defined only in the system and global state optimization cases).  For that case the algorithm can find a satisfactorily low cost for only the decoupled Hamiltonian where any unitary of the form $B=B_\sigma \otimes B_\epsilon$ will form a solution (although this does not imply that all solutions to $B$ must be tensor products; consider a SWAP gate for a trivial counter-example). Unsurprisingly, we do not get low cost solutions for the other Hamiltonians with this initialization approach; even though the cost is $\mathcal{O}(10^{-3})$ for the non-commuting central spin (second column), the solution does not generalize to arbitrary times (i.e. other than the ones on which it was trained). This non-generalizability  can be seen in Fig.~\ref{fig:violin-plot} of Appendix~\ref{sec:appendix-violin-plots}.  

In the second row, we additionally find generalizable solutions to the random tensor Hamiltonian. This is because, for $H=H_\sigma \otimes H_\epsilon$, the cost vanishes when $B$ is any tensor product unitary and $A_\sigma |\xi\rangle_\sigma = |s_i\rangle_\sigma$, where $|s_i\rangle_\sigma$ is an eigenstate of $H_s$. The non-commuting central spin has an even lower cost than in the top row because there is more freedom in choosing the global initial state but this is still not sufficient to generate a solution that generalizes to arbitrary times. 

Finally, the last row shows results for the case when we optimize over the global initial state. In this case, the algorithm is successful in finding generalizable solutions to all categories of Hamiltonians. There is a wide variety of solutions, each with a different physics interpretation, that are  available to the algorithm in this scenario; we elaborate more on the details of this solution space in Sec.~\ref{sec:analysis}. 

While Fig.~\ref{fig:algorithm_summary} only shows the most optimal solution, the detailed statistics for each class of Hamiltonian is discussed in Appendix~\ref{sec:appendix-violin-plots} and Fig.~\ref{fig:violin-plot} therein. The numerical results presented here are shown for a single qubit system and a two qubit environment ($n_s=1$, $n_e=2$, an $8$ dimensional Hilbert space).  We explored a variety of sizes up to $n_s=2$ and $n_e=4$ (a $64$ dimensional Hilbert space) and saw similar results. 
In the next section we demonstrate, using analytic forms of the destination Hamiltonian, how the low costs described here can be achieved by the algorithm. In all but one case, the analytic forms exist for arbitrary $n_s$ and $n_e$.  

\subsection{Analysis}
\label{sec:analysis}

\begin{figure}
    \centering
    \includegraphics[width=0.98\linewidth]{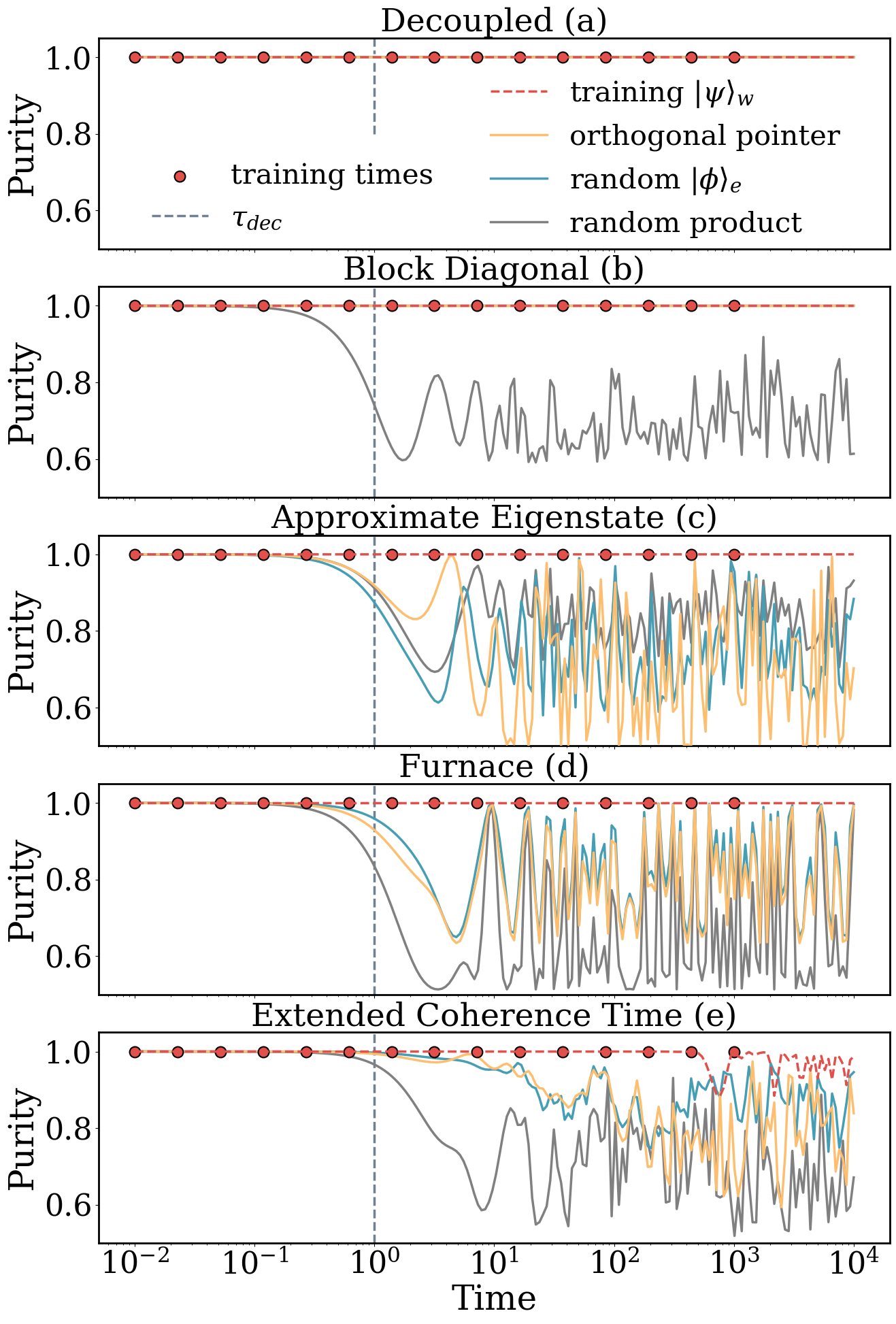}
    \caption{Time evolution of the purity of the reduced system density matrix  for each of the categories identified by our algorithm, as discussed in Sec.~\ref{sec:analysis}. When the purity is close to one, the training state is separable, i.e. $\ket{\psi}_w = |\chi\rangle_s|\phi_e\rangle$ in the destination factorization. Thus, we also show the effect on the purity resulting from substituting the system state for the orthogonal state, $|\chi\rangle_s \rightarrow |\chi'\rangle_s$, and of randomizing the environment state $|\phi\rangle_e$. While in the decoupled case all initial product states remain unentangled for arbitrary times, the other destinations only allow very particular states to stay unentangled beyond the decoherence time. The furnace case is interesting in that it admits a pointer state only for particular choices of the environment state, as elaborated on in Fig.~\ref{fig:furnace_details}. }
    \label{fig:Destination-Hams}
\end{figure}

There are many solutions that can satisfy the problem of identifying subsystems as we have posed it here by the vanishing of the cost given in Eq.~\eqref{eq:DisCostB}. In this section we offer a physical understanding of the different solutions we have found. We adopt the ``Heisenberg picture'' where we study the Hamiltonian in the new factorization.  

We have found the solutions can be grouped in five categories, each with its own characteristic behaviors. The \emph{Decoupled Hamiltonian} category is especially simple, but these solutions only exist for special cases. The \emph{Block Diagonal Hamiltonian} solutions exist even for completely random fiducial Hamiltonians, and correspond to perfectly stable static pointer states. The \emph{Approximate Eigenstate}, \emph{Furnace Hamiltonian} and \emph{Extended coherence time} each have more nuanced dynamics, and our understanding of these categories is a bit more phenomenological.  

Next, we discuss each category in turn, elaborating on the characteristics these solutions exhibit and also comment on their scalability to larger systems. Figure~\ref{fig:Destination-Hams}. shows the behaviors of specific solutions from each category, and is used to support the discussion which follows.  Which category of solution is found on a specific run of our algorithm involves an element of chance, as well as certain constraints which we also delineate below. 

\paragraph*{\textbf{  (a) Decoupled Hamiltonian.}} Here the Hamiltonian takes the form,
\begin{equation}
\label{eq:decoupled}
    H = H_s\otimes \mathbb{1}_e + \mathbb{1}_s \otimes H_e
\end{equation} so that the system and environment are totally decoupled.
In this case, the entropy of the initial state remains constant and an initial product state remains unentangled. 
However, this is a vanishingly rare solution that is not expected for random Hamiltonians. This can be understood heuristically by a simple degrees-of-freedom counting argument: decomposing a generic Hamiltonian in the Pauli basis gives a total of $4^{n_w}$ terms (where $n_w$ is the total number of qubits) which can be arranged in a matrix of Pauli coefficients of dimensions ($4^{n_s}$, $4^{n_e}$) in such a way that the first column and row correspond to the system and environment self-interaction terms respectively. Thus, there are a total of $4^{n_s}-1$ terms that contribute to the system self-interaction and $4^{n_e} - 1$ for the environment (the $-1$ occurs because there is a Pauli coefficient for the identity which only contributes a global phase and is not of physical significance). The rest of the $4^{n_s +n_e} - 4^{n_s} - 4^{n_e}+1$ terms contribute to the interaction Hamiltonian, $H_{\rm int}$. Thus, say for a 3-qubit world, expecting to find a factorization where the Hamiltonian is decoupled amounts to the expectation that the $45$ interaction coefficients can be regrouped into $20$ slots for the self-interaction terms~\footnote{ \label{footnote:eigen-constraints} Another way the decoupled form appears difficult to realize is that in the decoupled case the eigenvalues of $H_w$ are sums of pairs of the $n_s$ eigenvalues of $H_s$ and the $n_e$ eigenvalues of $H_e$. This creates rigid constraints on the $n_w$ eigenvalues of $H_w$, and it is not clear that such constraints can be generally realized. See~\cite{knutson2000honeycombs} for a related investigation.}.

A TPS that realizes a decoupled Hamiltonian cannot generally be found when starting with a random Hamiltonian, but our algorithm does find such a TPS  for the interacting central spin model. This case does pass the counting argument test since in the fiducial factorization, $H_{\rm int}$ contains only $n_e$ non-zero Pauli terms. We depict the result of evolving several product states in this decoupled factorization of the central spin model in Fig.~\ref{fig:Destination-Hams}; clearly they all remain product states throughout the evolution. 

    
The transformation of an interacting Hamiltonian to a decoupled one is familiar from bosonization of interacting field theories \cite{coleman1975quantum-bosonization, von1998bosonization-reformionization} or the Jordan-Wigner (JW) transformation. The latter is particularly pertinent for our central spin Hamiltonian since, for several classes of Hamiltonians that are quadratic in the spin operators, the JW transformation unitarily transforms the spin system to a set of free fermions \cite{lieb1961two-XY_JW, schultz1964two-lieb-Ising_JW}. We comment on the interpretation of such non-local transformations in Section~\ref{sec:relative-state}. Some authors have argued that factorizations with quasi-classical subsystems can be identified as those that minimize the interaction Hamiltonian \cite{tegmark2015consciousness, zanardi2022operational}. Ref.~\cite{mansuroglu2024quantum} presents a quantum algorithm optimized to search explicitly for factorizations in which the interaction term is minimized. While the decoupled case certainly passes that criterion, the requirement seems too restrictive. In the rest of this section, we show several destination Hamiltonians that admit pointer states, and thus a notion of quasi-classicality, despite having significant interactions between the subsystems. 
    
In looking at the second column of Fig.~\ref{fig:algorithm_summary}, one might wonder why, if such a decoupled Hamiltonian can be found for the central spin, the cost in the top row is considerably higher than in the bottom row. This is because, as mentioned in Sec.~\ref{sec:methodology}, our initial state is a product state in the \textit{fiducial} factorization where the Hamiltonian in Eq.~\eqref{eq:central-spin} has interactions. Though there exists a unitary such that $BHB^\dagger$ is decoupled (the destination TPS), the algorithm has no direct knowledge of this fact. Instead, it is tasked with minimizing the cost Eq.~\eqref{eq:DisCostB} (in the Schr\"odinger picture). But since $B\ket{\xi}_s\ket{\phi}_\epsilon$ is entangled, without initial state optimization the algorithm is unable to access a product initial state in the destination TPS, which is required to achieve a low cost for the decoupled destination Hamiltonian. When we include initial state optimization (last row of Fig.~\ref{fig:algorithm_summary}) then, as expected, the algorithm succeeds in finding this solution.  And not surprisingly the middle row, showing the partially optimized case, reflects a cost midway between the two extremes.\\ 

\paragraph*{\textbf{ (b) Block Diagonal Hamiltonian.}} For any Hamiltonian, there always exists a factorization, defined by the eigenstates of the Hamiltonian, that admits pointer states. Specifically, we show in Appendix~\ref{sec:appendix-BlockD} that for any Hamiltonian there always exists a factorization of the Hilbert space where the eigenstates are separable so that it may be written in the form
\begin{equation}\label{eq:destination-hamiltonian}
H = \sum_i |\chi_i\rangle \langle \chi_i|_s \otimes H_e^{(i)}
\end{equation}
It is straightforward to verify, as shown in Appendix~\ref{ap:construct}, that $|\chi_i\rangle_s$ are the pointer states and that a generic state will evolve into a state whose system density matrix is diagonal in the pointer basis, i.e. 
\begin{equation}    \label{eq:rhos_decoherence}
\rho_s \xrightarrow{U_t} \rho_s \approx \sum_i p_i(t)  |\chi_i\rangle \langle\chi_i|_s
\end{equation}
(for a sufficiently large and scrambling environment) as is characteristic of einselection. It is also manifest from the form of Eq.~\eqref{eq:destination-hamiltonian} that $H$ is block diagonal in the pointer basis and, furthermore, that the $|\chi_i\rangle_s$ are robust to entanglement \textit{independent of the choice of the environment state or the size or scrambling nature of the environment}.\footnote{The distinction we are making here is that the robustness addresses the evolution of a state where the system is initially in the pure state $|\chi_i\rangle_s$, and Eq.~\eqref{eq:rhos_decoherence} reflects the behavior of an arbitrary initial state.} This feature can be seen in Fig.~\ref{fig:Destination-Hams} where we show that randomizing the environment state does not degrade the purity of the reduced density matrix. We also note that the pointer states are stationary, while an arbitrary environment state will evolve under the evolution generated by the $H_e^{(i)}$ operators. All these properties, as well as a general recipe for finding this factorization for an arbitrary Hamiltonian, are proved in Appendix~\ref{sec:appendix-BlockD}. 

We note that such a mapping to a factorization where the energy eigenstates are separable has some similarity with the case of dynamical decoupling in superconducting qubits (see e.g. Appendix A in \cite{pokharel2018demonstration-lidar} for an overview). There, the ``logical'' qubits (what we call the ``destination'' TPS) correspond to the energy eigenstates of a world described by interacting ``physical'' qubits (what we have called the ``fiducial'' TPS). Defining the logical qubits in this way can mitigate certain sources of noise \cite{heunisch2023tunable-hartmann}.

In the decoherence literature authors often consider special Hamiltonians; such as those for which the system self-interaction commutes with the interaction, $[H_s \otimes 1_e, H_{\rm int}]=0$ or a ``quantum measurement limit'' where the Hamiltonian is dominated by a single interaction term $H\approx H_s \otimes H_e$. These are both special cases of the more general block-diagonalization procedure outlined above where the $H_e^{(j)}$ operators in Eq.~\eqref{eq:destination-hamiltonian} have specific relationships. For example, in the quantum measurement limit all the $H_e^{(i)}$ operators are the same up to an overall scaling, $H_e^{(j)} \propto H_e^{(i)}$. A particularly interesting sub-case of the block diagonal Hamiltonian is when there exists a decoherence-free subspace (DFS) (a topic relevant for quantum computing~\cite{Lidar_2003,quiroz2024dynamically-lidar-DFS}). In this factorization some, but not all, eigenvectors have vanishing entropy and, as in the quantum measurement limit, their corresponding $H_e^{(i)}$ operators are proportional to each other. Writing out the Hamiltonian in the global eigenbasis, we can interpret it as a sum of a decoupled (i.e. DFS) and an entangling part:
\begin{equation}
\begin{gathered}
    \label{eq:DFS}
    H = \sum_i^{n_w} |i\rangle\langle i|_w \lambda_i = \sum_{i \in C_{\rm DFS}} |i\rangle\langle i|_w \lambda_i + \sum_{i \in C_{\rm rem}}|i\rangle\langle i|_w \lambda_i \\
    \\
    = H_{\rm DFS} + H_{\rm rem}    \, .
\end{gathered}
\end{equation}
Here we have split up the eigendecomposition into a term containing only the zero entropy eigenvectors (labeled `DFS') and a term containing the remainder (labeled `rem'). Thus, the disjoint set of indices satisfy the condition $C_{\rm DFS} \cup C_{\rm rem} = \{1,...,d_w\}$ and the operators $H_{\rm DFS}$ and $H_{\rm rem}$ are partial-rank. Now the reduced density matrix of any state composed of only the zero entropy eigenstates,
\begin{equation}
    \label{eq:DFS_state}
    \ket{\psi} = \sum_{i \in C_{\rm DFS}} \alpha_i |i\rangle_w
\end{equation}
will have \textit{constant} purity (which may be vanishing for certain choices of $\alpha_i$) and have evolution generated under $H_{\rm DFS}$. 

Such a DFS solution lies in the intermediate regime of the general block diagonal form in Eq.~\eqref{eq:destination-hamiltonian} and the decoupled Hamiltonian in Eq.~\eqref{eq:decoupled}. Since $H_{\rm DFS}$ is subject to the same constraints as those for the (fully) decoupled Hamiltonian (see footnote~\ref{footnote:eigen-constraints}), we do not expect to find such a factorization for random Hamiltonians. However, the central spin model admits both the decoupled and the block diagonal destinations, and in that case our algorithm is indeed able to find this intermediate DFS solution. \\

\begin{figure}
\centering
\includegraphics[width=0.43\textwidth]{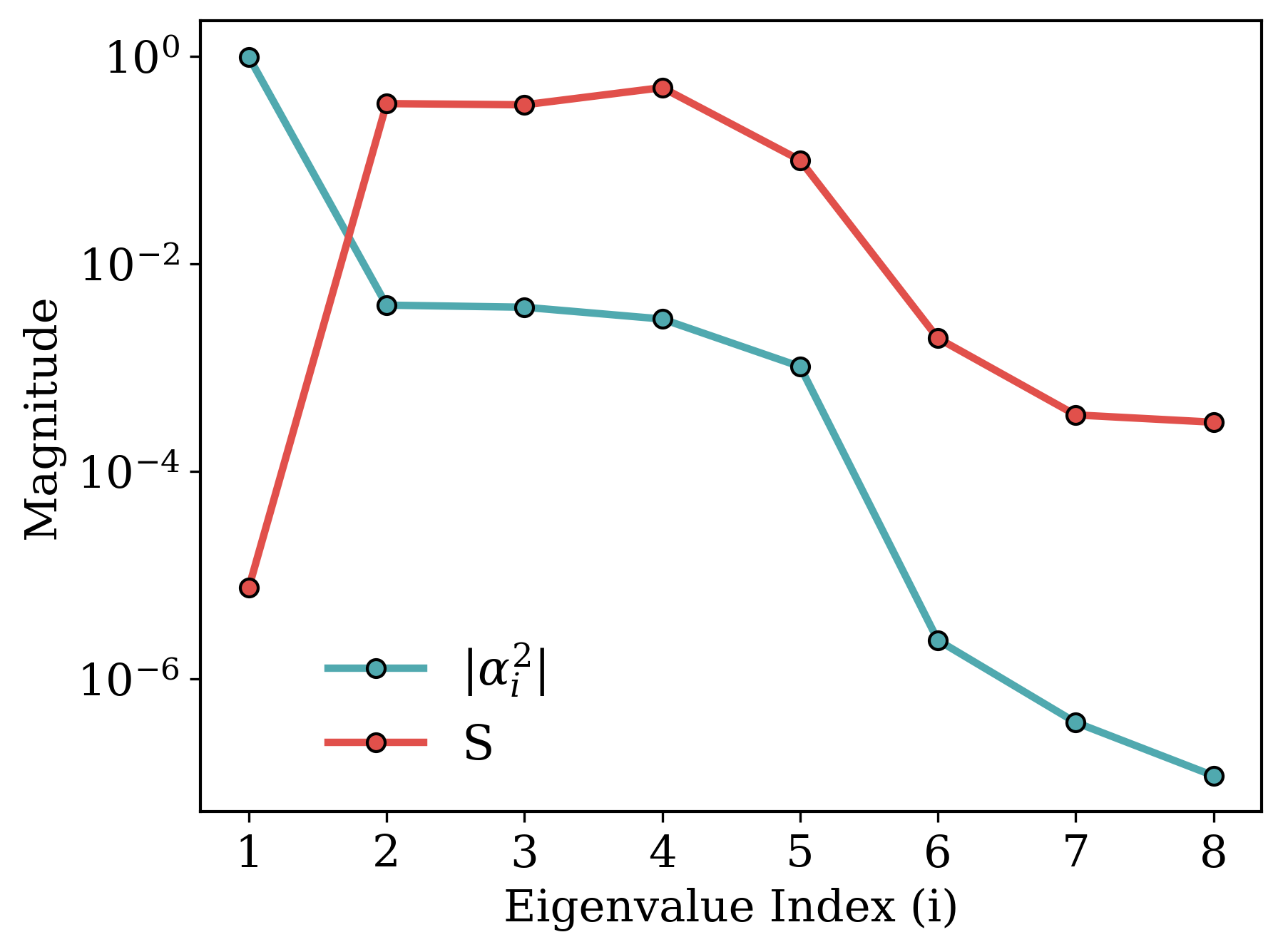}
\caption{The decomposition of $\ket{\psi}_w$ in the energy eigenbasis for the case of the `approximate eigenstate'' solutions discussed in Sec.~\ref{sec:analysis} (see Eq.~\eqref{eq:approx-eig-psi}) is shown in blue along with their corresponding linear entropy (red) of each energy eigenstate (defined in Eq.~\eqref{eq:lin-entropy}). These solutions are dominated by a single low entropy energy eigenstate, although high entropy eigenstates contribute to a small degree. The plot is arranged in order of decreasing $|\alpha_i|$.}
\label{fig:approx-eigenstate-decomp}
\end{figure}

\paragraph*{\textbf{(c) Approximate Eigenstates.} } This class of solutions is particularly interesting because it exhibits dynamic pointer states, though it fails to provide a complete basis of such states. Here, there is a dominant contribution to $\ket{\psi}_w$ from a low entropy energy eigenstate and smaller but, crucially, non-negligible contributions from high entropy eigenstates. More formally, in the language of Appendix~\ref{sec:appendix-BlockD}, the world energy eigenstates are of the form,
\begin{equation}
\begin{aligned}        
&\ket{1}_w = \ket{1}_s \ket{1}_e\\    
&\ket{i}_w = \sum_{j}^{d_s} \beta_{j}^{(i)} \ket{j^{(i)} }_s \ket{j^{(i)}}_e
\end{aligned}
\label{eq:approx-eig-psi1}
\end{equation}
where $\ket{1}_w$ has zero entropy, and we take the 
Schmidt coefficients $\beta_{j}^{(i)} $ to be generally nonzero for $j> 1$ $\forall i \in \{2,..,d_w \}$, so the other energy eigenstates have nonzero entropy.  (for convenience we choose to label the low entropy eigenstate $\ket{1}_w$). And the global state is of the form, 
\begin{equation}  
\begin{gathered}  
\ket{\psi}_w = \sum_{i=1}^{d_w} \alpha_i |i\rangle_w \\
\textrm{such that } \alpha_1\gg\ \alpha_j \textrm{   for   } j>1.
\end{gathered}
\label{eq:approx-eig-psi}
\end{equation}
This decomposition of $\ket{\psi}_w$ into eigenstates $\ket{i}_w$ and their corresponding linear entropy of entanglement, Eq.~\eqref{eq:lin-entropy}, is shown in Fig.~\ref{fig:approx-eigenstate-decomp} for a world with one system and two environment qubits~\footnote{We note that the correlation between the entropy and $|\alpha_i|^2$ for $i>2$ seems typical of these `approximate eigenstate' solutions but that it is likely a feature of the search algorithm and not of inherent physical significance. We have checked numerically that, as long as the condition  $\alpha_1 \gg \alpha_{j>1}$ is satisfied, there is considerable freedom in distributing the weights amongst the other higher entropy eigenstates.}. In Fig.~\ref{fig:Destination-Hams}~c. we show the evolution of the purity of the reduced system density matrix of such a state (labeled `training $\ket{\psi}_w$') compared to other states.
    
The fact that the contribution to $\ket{\psi}_w$ from the high entropy eigenstates is small, but not negligible, is crucial as it leads to a pure $\rho_s$ for times several orders of magnitude greater than the decoherence time (see Fig.~\ref{fig:Destination-Hams}~c.), while allowing the state to be dynamic. These two behaviors resonate with our classical experience of subsystems, and cannot be achieved by the single low entropy eigenstate on its own. 

It appears, at first, that such solutions are exactly the ones that the author of \cite{tegmark2015consciousness} is in search of, i.e. those that have a high degree of predictability (as evidenced by the purity of the system state) as well as ``autonomy'' (that their dynamics are not trivial). We note that while this class of solutions satisfactorily minimizes our cost function (Eq.~\eqref{eq:DisCostB}), they do not yield a multi-dimensional distinguishable basis of pointer states of the form in Eq.~\eqref{eq:approx-eig-psi}. 
This should be contrasted with the block diagonal TPS (c.f. part (b)) which has a complete set of pointer basis states so that the system density matrix follows Eq.~\eqref{eq:rhos_decoherence}. However, in that case, the pointer states have trivial dynamics. 
While requiring a complete pointer basis seems important for standard discussions of einselection, singling out a special low entropy state (i.e. Eq.~\eqref{eq:approx-eig-psi1}) might be relevant for discussions of cosmological initial conditions. \\ 

\paragraph*{\textbf{(d) Furnace Hamiltonian.} }
\label{subsec:Furnace}
In this scenario there exist pointer states when the environment state exists in a {\it particular} subspace of $\mathcal{H}_e$. Specifically, suppose 
\begin{equation}
\label{eq:furnace}
H = \sum_i^{d_e} H_s^{(i)}\otimes|i\rangle\langle i|_e
\end{equation}
and pick, for simplicity, two particular but arbitrary indices $p$, $q$ such that $[H_s^{p}, H_s^{q}] = 0$ . Then, any initial state of the form $|\psi\rangle = |s_i\rangle (\alpha |p\rangle_e + \beta |q\rangle_e)$ (where $|s_i\rangle$ is a mutual eigenstate of $H_s^{p}$ and $H_s^{q}$) will remain unentangled. Of course, while we picked just two commuting $H_s^{(i)}$ terms for simplicity, the argument is quite general; all that is needed is for the environment state to project the total state into a subspace that admits a pointer state solution.

\begin{figure}
    \centering
    \includegraphics[width=0.95\linewidth]{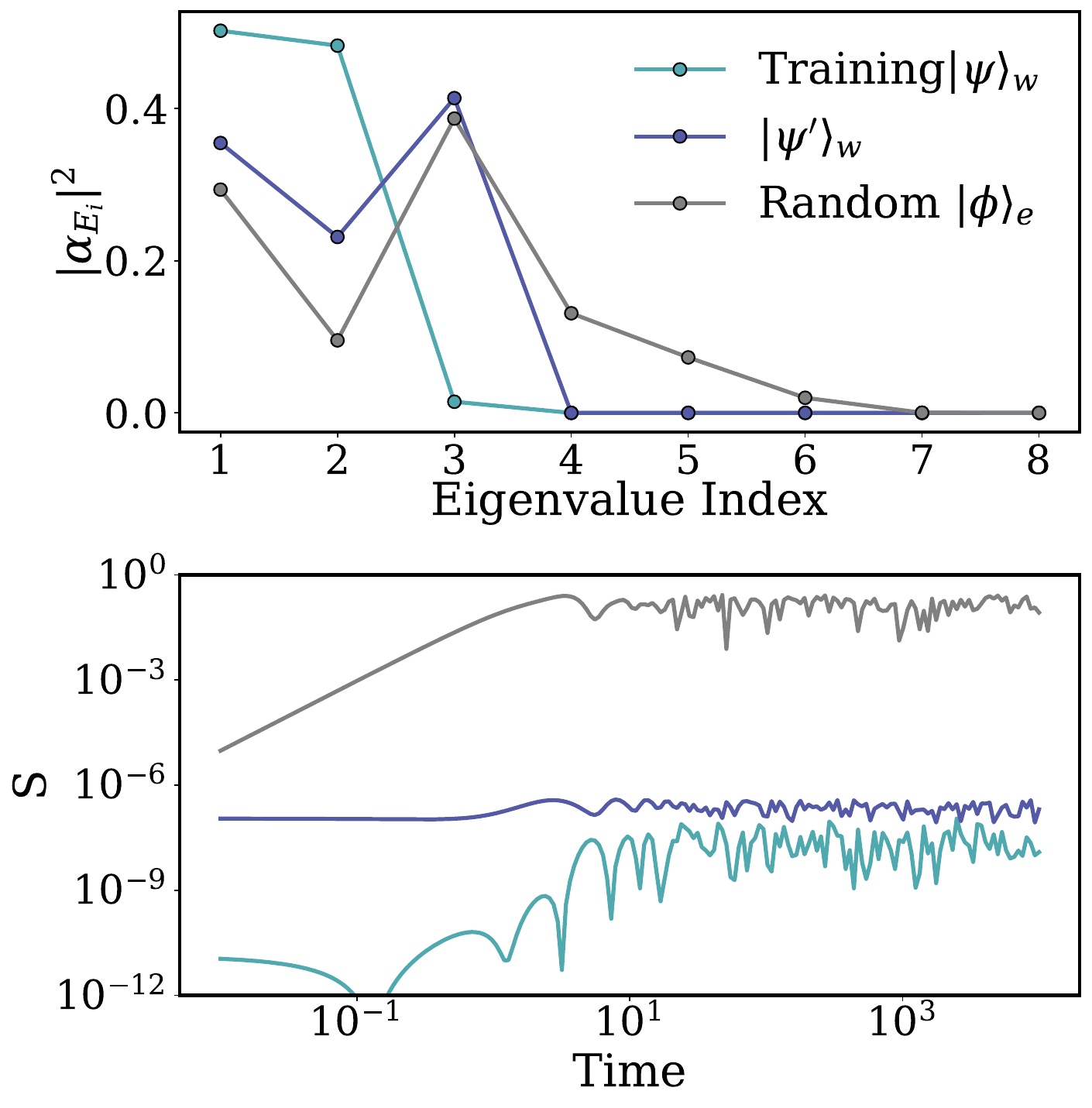}
    \caption{In the furnace Hamiltonian, a pointer state for the system exists only if the environment is in a particular subspace of $\mathcal{H}_e$ (unlike the Block diagonal case, c.f. Fig.~\ref{fig:Destination-Hams}b, where the pointer state is indifferent to the choice of $|\phi\rangle_e$). Here we depict this phenomenon by perturbing the trained state $|\psi\rangle_w$ (turquoise) such that the new state $|\psi'\rangle_w$ (purple)  has support from the same energy eigenvectors as those that contribute to the trained state (top panel). The bottom panel verifies that such a state indeed remains a product state at late times. Note that $|\psi\rangle$ and $|\psi'\rangle$ share the same system state in $\mathcal{H}_s$ (up to an overall phase). We show, for comparison, another state where $|\phi\rangle_e$ is randomized leading to a large increase in the linear entropy in the typical decoherence time (the furnace ``incinerates'' in the analogy of the text in Section~\ref{sec:analysis}(d)).  }
    \label{fig:furnace_details}
\end{figure}

The form of the Hamiltonian in Eq.~\eqref{eq:furnace} also makes it clear that, for a fixed system size, the larger the environment the more such solutions will exist. This is because there are $d_e$ terms and, if each of the $H_s^{(i)}$ operators is generated randomly, the likelihood that any two operators will commute (or approximately commute) increases. As shown in Fig.~\ref{fig:Destination-Hams}d, clearly the trained state remains unentangled throughout the evolution while the other states, including the state made from randomizing the environment state, generally do not. Although, in this figure, the `furnace' solution appears to resemble the `approximate eigenstate' solution, the distinguishing feature can be seen in Fig.~\ref{fig:furnace_details} which shows that the purity of the reduced density matrix of the trained state is robust to arbitrary changes within a particular subspace spanned by a subset of the energy eigenstates. 
     
Intuitively, this solution can be understood in the following physical sense: consider a pendulum in the lab that undergoes constant environment monitoring and thus stays in the pointer state as it oscillates (as is expected of a subsystem). But now the lab assistant turns up the temperature of the room, as in a furnace, so that the pendulum incinerates and its atoms are dispersed amongst the various degrees of freedom available in the room. Here, though the Hamiltonian has remained the same, changing the environment state has drastically altered the description of the system to one that no longer has a well-defined classical subsystem.\\ 
     
\paragraph*{\textbf{(e) Extended coherence time.}} These solutions sustain a moderately high purity for $1-2$ orders of magnitude longer than the typical decoherence time. From the algorithm's perspective such solutions can be thought of as a result of poor training or poor generalizability. In the former case, the algorithm finds a combination of initial state and factorization such that the solution only \textit{roughly} approximates one of the other ``perfect'' solutions above. However, the convergence to such a solution is slow and the algorithm hits a time-wall before it can get sufficiently close to such a solution. Poor generalizability occurs when the algorithm is able to sufficiently minimize the cost for the training steps, but that the solution does not generalize well beyond those finite times. In this case, the situation resembles the purity plots shown in the left panel of Fig.~\ref{fig:training_times}. As mentioned in Section~\ref{sec:methodology}, since we take our maximum training time to be $\approx 10^2 t_{\rm dec}$, it makes sense that the failure to generalize manifests beyond this domain. 

Such `extended coherence time' solutions also occur in the case of fixed input state optimization (i.e. $A=\mathbb{1}$). This is because, in fixing $A$, the algorithm loses control over the degrees of freedom required to sufficiently minimize the cost and to ensure that the purity of the reduced density matrix remains high for times much greater than $t_{\rm dec}$. We find these solutions for all classes of Hamiltonian and they may be interpreted as short-lived subsystems. That is, as a set of dynamics which, \textit{a priori}, appears highly entangling but can be viewed in another factorization as temporarily pertaining to subsystems. Ultimately, we lack an analytic understanding of these solutions and caution from reading too much into the numerics shown here. Firstly, the purities obtained in this case are substantially lower than in the case where we optimise over the initial state (see the right panel of Fig.~\ref{fig:lin_entropy_all_H}). Secondly, it is unclear how these solutions generalize to larger systems. On the other hand, these solutions might most closely reflect the fate of all subsystems in a dark energy dominated universe, which could ultimately end up as a thermalized de Sitter space. 

To summarize, we have found five different categories of solution to our optimization problem, each with its distinct set of associated behaviors. The block diagonal destination Hamiltonian is \textit{always} available as a TPS that supports einselection, even for random Hamiltonians. Thus, one need not construct special Hamiltonians to demonstrate decoherence.  
The decoupled case can be realized for certain structured Hamiltonians. The other three cases are intriguing in their own right and each offers its own set of open questions (such as how important is it to have a \emph{complete} pointer basis, and what patterns of energy eigenstate entropies are conducive to classical dynamics). Perhaps most intriguing are cases where these different destination Hamiltonians can be found for the same set of eigenvalues.The implication is that different types of subsystem coexist, at least formally, within the same Hamiltonian.  We next explore some implications of this result for quantum foundations. 

\section{Many \textit{more} worlds}
\label{sec:relative-state}
Our results appear to have provocative implications for certain interpretations of quantum mechanics. We will not make value judgments on the interpretations themselves but rather point out that the presence of different TPS's that coexist, describing alternative semiclassical subsystems of the same larger system leads to an ambiguity in the global basis with respect to which the branches of the wavefunction may be identified with classical reality. 

Everett’s relative state interpretation of quantum mechanics \cite{everett1956Thesis,everett1957relative, wheeler1957assessment} was meant to address the system-observer dichotomy, where the observer/apparatus is treated as a classical object (and many textbooks continue to present this dichotomous exposition). The relative state interpretation, and its various extensions in de Witt’s many-worlds interpretation \cite{dewitt1973many-worlds} or Zurek’s existential \cite{zurek1998decoherence-existential-interpretation-rough-guide} interpretation, posits a global wave function that evolves unitarily and definite perceived outcomes result from the state of the observer becoming correlated with the state of the system. It is in this sense that the relative state interpretation may be thought of as a literal reading of unitary quantum mechanics. For concreteness, consider a system in the state 
\begin{equation}
 |\psi\rangle_s =  \frac{1}{\sqrt{2}}(|0\rangle + |1 \rangle)  
 \label{eq:prefered-basis-start}
\end{equation} that interacts with an apparatus such that  
\begin{equation}
 \frac{1}{\sqrt{2}}(|0\rangle + |1 \rangle) |a_r\rangle \xrightarrow{U_t} |\Psi_f \rangle = \frac{1}{\sqrt{2}} (|0\rangle|a_0\rangle+|1\rangle|a_1\rangle) 
\label{eq:prefered-basis}
\end{equation}
where $|a_r\rangle$ is the ``ready'' state of the apparatus. The relative state interpretation posits that no definite state can be ascribed to the system or the apparatus but, instead, the state of the apparatus (or the observer), say $|a_1 \rangle$, should be understood as being correlated with the state $|1\rangle$ of the system. Then, each of the two terms of Eq.~\eqref{eq:prefered-basis} correspond to the two ``branches'' of the wave-function. 

However, the relative interpretation then faces the so-called `preferred basis' problem. Namely, why should the branches be described by the particular choice of basis used in Eq.~\eqref{eq:prefered-basis}? Consider instead the system and apparatus bases,
\begin{equation*}
    |\pm\rangle_s = \frac{1}{\sqrt{2}}(|0\rangle \pm |1\rangle)
    \textrm{  and  }
|a_\pm\rangle = \frac{1}{\sqrt{2}}(|a_0\rangle \pm |a_1\rangle)  ,
\end{equation*}
so that the final global state in Eq.~\eqref{eq:prefered-basis} takes the form,
\begin{equation}
    |\Psi_f\rangle = \frac{1}{\sqrt{2}} (|+\rangle |a_+\rangle + |-\rangle |a_-\rangle) \; .
    \label{eq:prefered-basis-2}
\end{equation}
This ambiguity in the choice of basis between Eq.~\eqref{eq:prefered-basis} and Eq.~\eqref{eq:prefered-basis-2} poses the questions:  what exactly has the apparatus ``measured''? In the language of the many worlds interpretation, which basis (that of Eq.~\eqref{eq:prefered-basis} or \eqref{eq:prefered-basis-2}) describes the post-measurement world?\footnote{The fact that the two terms in Eq.~\eqref{eq:prefered-basis-start} have equal coefficients is a choice that simplifies our mathematical expressions, but there are equivalent issues for the general case.} 

Einselection resolves this preferred basis problem \cite{zurek2003decoherence-and-the-quantum-origins} by considering the interaction of the apparatus with the environment. Depending on the details of the Hamiltonian, this apparatus-environment action singles out a set of preferred states so that, say, the $\{a_0,a_1\}$ basis are pointer states that remains robust to entanglement with the environment. The state of the system should be inferred relative to this basis.  
In this manner, einselection provides a way to dynamically select the ontologically significant branches of the wavefunction for relative state interpretations. 

However, the presence of pointer states is a crucial assumption in previous work on einselection
and requires specially constructed Hamiltonians (whether through the predictability sieve strategy \cite{zurek1993preferred-predictability-sieve} or by virtue of vanishing commutators in the self and interaction  Hamiltonians \cite{zurek1981pointer}). In particular, it requires one to specify the individual subsystems that the Hamiltonian describes. 

In this paper, we set out to answer the question of whether the division into subsystems can be derived, as opposed to assumed, from the spectrum of the Hamiltonian\footnote{This is similar in spirit to D. Deutsch's ``interpretation basis'' approach \cite{deutsch1985quantum-interpretation-basis} where he proposes a resolution to the preferred basis problem that can, in turn, be used to address the issue of factorization into subsystems (see Eq. 50 and surrounding discussion therein). However, see Ref.~\cite{foster1988recent-interpretation-basis} for crucial arguments against the interpretational basis approach. }. In Section~\ref{sec:analysis}, we have shown that there always exists at least one such factorization -- that in which the global eigenbasis is separable, that admits pointer states and hence can exhibit einselection (in the sense of Eq. \ref{eq:prefered-basis} above). Moreover, for ``real-world'' Hamiltonians, several other choices also exist. But, this introduces a new issue: the block diagonal factorization is available \textit{in addition} to the other factorizations that also support einselection (such as the coherent state wavepackets of the ACL model discussed in Section~\ref{sec:what-is-a-subsytem}). What then favors one factorization over the other?  

Our findings in this paper thus highlight a “global” analog of the preferred basis problem: if multiple TPSs admit consistent histories why should the world branch in one factorization over the other? Why should the physicist’s choice of \textit{global} basis in writing down the state  determine the branches of the worlds? Moreover, the resulting ambiguity is in a sense more unsettling than the preferred basis ambiguity of the system subspace. The various choices of TPS describe many ``realms'': the set of worlds described by the fiducial factorization have a different ontology than the set of worlds in the realm described by the destination factorization.
The harmonic oscillator of the ACL is far from an SHO in a different system-environment split. And while both realms share the same global energy spectrum, the scene described by a central spin surrounded by environment qubits in an external magnetic field is radically different than the fermionic degrees of freedom with hopping terms obtained by taking a Jordan-Wigner transform.

While in this paper we have focused on the case of bipartite factorizations, this new preferred basis problem (in contrast to the resolution of the original preferred basis problem) looks unlikely to be resolved by expanding to tripartite/multipartite splits. 
Namely, the case of multipartite subsystems may further add to this diversity of the realms, especially if one considers $\dim{(\mathcal{H}_w)}$ to be highly factorizable as appears to be the case for our Universe \cite{eakins2002factorization-microsingularity}. 

Thus, this ontological ambiguity between the realms is unsettling, and some might want to refer to a particular “physical” factorization as describing “real” subsystems, and the other factorizations describing  subsystems which are “virtual” or in some sense less “real”, we reserve judgment on the “reality” of eccentric realms \footnote{Coleman poses the same issue in the context of bosonization \cite{coleman1975quantum-bosonization}, ``One of the most striking features of the results established here is that a theory which is `obviously' a theory of fermions is equivalent to a theory of bosons... That these theories are equivalent in any sense seems too preposterous to believe...''  } .  
One might argue that the “physical” factorizations have some notion of locality either in the pointer state as in \cite{piazza2010glimmers, carroll2021quantum-singh}, at the level of the Hamiltonian \cite{Cotler:2017abq}, or in the set of observables \cite{zanardi2004quantum-lloyd-lidar}. While there is some merit to this argument, we remain curious if the ``local'' factorizations are in some deeper sense preferred over the other ones and whether one can derive this preference from something fundamental rather than inflicting this classical prejudice onto the Hilbert space. Assuming locality is principally not much different than assuming a particular factorization of the Hamiltonian as is done implicitly in most formulations, and explicitly in Wallace's three postulates \cite[Chapter~1]{wallace2012emergent-multiverse} of quantum mechanics.

Similar prejudices enter into arguments that try to minimize $H_{\rm int}$ in an effort to look for a preferred factorization \cite{tegmark2015consciousness, zanardi2022operational}.  But there is nothing in the einselection formalism (or the relative state interpretations) that presupposes a TPS that minimizes the interaction Hamiltonian.\footnote{A general discussion of einselection behaviors under different interaction strengths appears in~\cite{schlosshauer2007decoherence} and several cases are illustrated in concrete terms for the ACL model in~\cite{albrecht2023adaptedcalmodel}.} Attempts to search for such a preferred TPS by minimizing the interaction strength are misguided because important physical cases are described by a dominant $H_{\rm int}$ term and in those cases einselection occurs not in spite of but \textit{because} of the strong interaction (see \textit{Quantum Measurement Limit} in  Section~\ref{subsec:SR-Hamiltonians-considered}). 

While others have put forth ``in-principle''  algorithms to search for quasi-classical factorizations \cite{deutsch1985quantum-interpretation-basis, carroll2021quantum-singh}, we have been successful in practically implementing an algorithm to explore the many realms that coexist in worlds with a given global energy spectrum. What remains missing is some measure on the space of these realms serving much the same purpose as that served by the predictability sieve for the preferred basis problem, or by the Born rule for the many world branches (to suppress the ``maverick'' worlds \cite{dewitt1970quantum-Physics-Today}). As we explain in Appendix~\ref{sec:appendix-violin-plots}, one such measure is induced by our algorithm based on the minimization of the cost Eq.~\eqref{eq:DisCostB}. However, it is unlikely that such a measure is of any physical significance.


The discussion in this section has taken a predominantly ``cosmological'' point of view, where care is taken not to refer to an external observer.  From that point of view one can wonder if there ought to be some physical principle that allows us to choose a ``correct'' realm from the many offered up by the formalism. 

We note that one could instead take an ``engineering'' perspective and think of the systems we explore in this paper as actual sets of qubits sitting in the laboratory.  In that case it is clear that each of the coexisting einselecting realms we've identified are equally physical, and could in fact be relevant to us if we are able to develop probes (from our larger world or just from another laboratory system of qubits) that can interact with the particular subsystem in question.  We are intrigued by the sharp contrast this engineering perspective makes in comparison to the cosmological one. 

Let us end this section by commenting on potential links between our findings here and investigations elsewhere. Firstly, the abundance of coexisting realms, and the absence of a physical measure to discriminate among them appears to be related to issues that have been explored in the context of the consistent histories formalism, where alternative sets of complete projectors lead to different coexisting narratives~\cite{Albrecht:1992rs,Albrecht:1992uc,Dowker:1994ac, Brun_Hartle_1999, wallace2002worlds-consistent-histories} (``alternate facts'', in the language of~\cite{Albrecht:2021sic}).  However, those discussions typically do not explore as wide a range of tensor product structures as we consider in this paper. The need for more information than is contained in a Hamiltonian to determine a unique tensor product structure has also been investigated from a mathematical perspective in~\cite{Stoica:2021bub,Stoica:2021rqi,Stoica:2024nhh}. Finally, we note that there are also some parallels with discussions of the ``clock ambiguity'' in the context of quantum gravity~\cite{Albrecht:2007mm}.

\section{Conclusions}
\label{sec:conclusions}
In this work we have addressed the question: \textit{Given a quantum world, determined by its energy spectrum, can we find factorizations of the global Hilbert space such that certain states of the system are robust to entanglement?}   The dynamical selection of such robust (pointer) states, known as einselection, has a key role in the emergence of classical from quantum allowing systems in pointer states to be tracked classically without consideration of various quantum processes. 
We have explored this question both numerically and analytically. On the numerical side, we introduced an algorithm to minimize a cost function that is sensitive to the purity of the evolving subsystem quantum state. This gives us an operational tool to identify pointer states which we use to numerically explore systems comprised of several qubits. Complementary to these numerical explorations, we have also found several analytic forms of Hamiltonian describing subsystems with quasi-classical properties. We find these analytic forms interesting in their own right, and also very helpful for interpreting the different solutions found by the algorithm. 

For a given global Hamiltonian (selected from one of four classes) we utilized a variety of optimization schemes for exploring possible tensor product structures and initial states. We have found that in all cases multiple tensor product structures can be found describing coexisting ``realms'' that exhibit features of classicality. This was so even for cases with randomly generated energy eigenvalues, and cases with a more structured energy spectrum allowed for a greater variety of classical behaviors. We note that the interactions in these quasi-classical realms need not be weak or local.

We have explored a variety of implications of our results.  From a foundational point of view, our results compound interpretational questions that arise when multiple classical worlds coexist within a single quantum description of the Universe. At a more practical level, our results also hint towards new strategies for developing noise robust quantum technologies. The systems we have studied may be viewed as real laboratory objects, with each of the realms having a physical existence which could be examined if suitable probes could be developed.  We call out in particular the possible relevance of our results for the engineering of decoherence-free subspaces. 

The numerical results presented here are shown for a single qubit system and a two qubit environment ($n_s=1$, $n_e=2$).  We explored a variety of sizes up to $n_s=2$ and $n_e=4$ and saw similar results. The analytical forms of the ``destination'' Hamiltonians that we derived in Section~\ref{sec:analysis} exist for arbitrary $n_s$ and $n_e$ (with the exception of case (e) whose scaling behavior is unclear). For example, the block diagonal case is always available and would, in theory, give a vanishing cost function for any values of $n_s$ and $n_e$. Our algorithm was helpful in discovering what kinds of solutions can exist.  The analytic forms were developed to understand the various solutions that the algorithm returned to us, and we view the analytic work as a key part of our results. 


It would be interesting to extend this exploration to larger values of $n_s$ and $n_e$ and see what other kinds of solutions appear.   Another important extension would involve expanding our treatment to numbers of subsystems larger than two. This is relevant to the Quantum Darwinism phenomena exhibited in our classical world~\cite{zurek2009quantum-darwinism} and also for constructing the sorts of probes we have mentioned in the context of applications to quantum engineering. Furthermore, while we have demonstrated the existence of several ``destinations'' for some classes of Hamiltonians, an intriguing open question remains regarding the fundamental structures of the spectrum to which the realms owe their existence.

\edits{Finally, in this work we have focused on einselection as the tracer for the quantum to classical transition. In doing so, we have used \textit{predictability} as the guiding principle for identifying quasi-classical subsystems. As briefly discussed in Sections~\ref{sec:Intro} and~\ref{sec:methodology}, einselection is but one aspect of classicality. We note that several authors (e.g.~\cite{Halliwell_1998_decoherent_hydrodynamics,Brun_Hartle_1999, Halliwell1999_decoherent_coarse_graining, Gell-Mann:2006ewf}) have considered the role of coarse graining, locally conserved quantities, and fast versus slow moving modes in determining which physical subsystems behave quasi-classically.
We have not studied in detail how these broader ideas might relate to the ``realms'' we have identified. We regard these connections as an interesting topic for future work. We do note that behaviors we have identified based on the properties of the global energy eigenstates extend in straightforward ways to arbitrarily large systems and we expect that includes systems which admit these other analyses of classical phenomena (i.e. those in terms of coarse graining etc.).  }


We have not attached a lot of conceptual importance to the fine details of our algorithm. In Sec.~\ref{sec:relative-state} we considered foundational questions which, perhaps in a cosmological context, demand some sort of measure with which to give weight to the different coexisting realms.  While our algorithm does operationally provide a measure, we have no reason to think it offers one that is relevant to such questions.  On the other hand,  we are curious if our algorithm (or a related one) might be relevant for practical applications where probes of the qubit systems need to be suitably engineered.  In principle the algorithm might reflect some aspects of the engineering process. 

\section{Acknowledgements}
Adil, Albrecht and Sornborger acknowledge support from the U.S. Department of Energy, Office of Science, Office of High Energy Physics QuantISED program under Contract No. KA2401032. ZH acknowledges support from the Sandoz Family Foundation-Monique de Meuron program for Academic Promotion.

A. Adil would like to thank the organizers and participants of the ``Foundations Under the Midnight Sun 2023'' workshop for several engaging discussions. We thank Lukasz Cincio for help with implementing the numerical algorithm and Michael Hartmann for helpful conversations about the application to superconducting qubits. We also thank M.~Lyans for an editorial suggestion, P.~Coles for helpful discussions at early stages of this work, and F.~Suzuki, A.~Touil, B.~Yan and W.~Zurek for comments on our original preprint. 

\bibliography{ref.bib}

@article{Albrecht_1994,
   title={Some Remarks on Quantum Coherence},
   volume={41},
   ISSN={1362-3044},
   url={http://dx.doi.org/10.1080/09500349414552311},
   DOI={10.1080/09500349414552311},
   number={12},
   journal={Journal of Modern Optics},
   publisher={Informa UK Limited},
   author={Albrecht, Andreas},
   year={1994},
   month=dec, pages={2467–2482} }

@article{Halliwell1999_decoherent_coarse_graining,
  title = {Decoherent Histories and the Emergent Classicality of Local Densities},
  author = {Halliwell, J. J.},
  journal = {Phys. Rev. Lett.},
  volume = {83},
  issue = {13},
  pages = {2481--2485},
  numpages = {0},
  year = {1999},
  month = {Sep},
  publisher = {American Physical Society},
  doi = {10.1103/PhysRevLett.83.2481},
  url = {https://link.aps.org/doi/10.1103/PhysRevLett.83.2481}
}

@article{Halliwell_1998_decoherent_hydrodynamics,
   title={Decoherent histories and hydrodynamic equations},
   volume={58},
   ISSN={1089-4918},
   url={http://dx.doi.org/10.1103/PhysRevD.58.105015},
   DOI={10.1103/physrevd.58.105015},
   number={10},
   journal={Physical Review D},
   publisher={American Physical Society (APS)},
   author={Halliwell, J. J.},
   year={1998},
   month=oct }

@article{Brun_Hartle_1999,
   title={Classical dynamics of the quantum harmonic chain},
   volume={60},
   ISSN={1089-4918},
   url={http://dx.doi.org/10.1103/PhysRevD.60.123503},
   DOI={10.1103/physrevd.60.123503},
   number={12},
   journal={Physical Review D},
   publisher={American Physical Society (APS)},
   author={Brun, Todd A. and Hartle, James B.},
   year={1999},
   month=nov }

@article{Omnes:1988fv,
    author = "Omnes, Roland",
    title = "{Logical Reformulation of Quantum Mechanics. 4. Projectors In Semiclassical Physics}",
    reportNumber = "LPTHE-ORSAY-88-60",
    doi = "10.1007/BF01023649",
    journal = "J. Statist. Phys.",
    volume = "57",
    pages = "356--382",
    year = "1989"
}

@article{Omnes:1988ej,
    author = "Omnes, Roland",
    title = "{Logical Reformulation of Quantum Mechanics. 3. Classical Limit and Irreversibility}",
    reportNumber = "LPTHE-ORSAY-88-12-3, LPTHE-ORSAY-87-47",
    doi = "10.1007/BF01014232",
    journal = "J. Statist. Phys.",
    volume = "53",
    pages = "957--975",
    year = "1988"
}

@article{Omnes:1988em,
    author = "Omnes, Roland",
    title = "{Logical Reformulation of Quantum Mechanics. 2. Interferences and the Einstein-Podolsky-Rosen Experiment}",
    reportNumber = "LPTHE-ORSAY-88-12-2, LPTHE-ORSAY-87-46",
    doi = "10.1007/BF01014231",
    journal = "J. Statist. Phys.",
    volume = "53",
    pages = "933--955",
    year = "1988"
}

@article{Omnes:1988ek,
    author = "Omnes, Roland",
    title = "{Logical Reformulation of Quantum Mechanics. 1. Foundations}",
    reportNumber = "LPTHE-ORSAY-88-12-1, LPTHE-ORSAY-87-45",
    doi = "10.1007/BF01014230",
    journal = "J. Statist. Phys.",
    volume = "53",
    pages = "893--932",
    year = "1988"
}

@article{zurek1993preferred-predictability-sieve,
  title={Preferred states, predictability, classicality and the environment-induced decoherence},
  author={Zurek, Wojciech H},
  journal={Progress of Theoretical Physics},
  volume={89},
  number={2},
  pages={281--312},
  year={1993},
  publisher={Oxford University Press}
}

@article{Gell-Mann:2006ewf,
    author = "Gell-Mann, Murray and Hartle, James",
    title = "{Quasiclassical coarse graining and thermodynamic entropy}",
    eprint = "quant-ph/0609190",
    archivePrefix = "arXiv",
    doi = "10.1103/PhysRevA.76.022104",
    journal = "Phys. Rev. A",
    volume = "76",
    pages = "022104",
    year = "2007"
}

@article{Gell-Mann:2011rif,
    author = "Gell-Mann, Murray and Hartle, James B.",
    title = "{Decoherent Histories Quantum Mechanics with One `Real' Fine-Grained History}",
    eprint = "1106.0767",
    archivePrefix = "arXiv",
    primaryClass = "quant-ph",
    doi = "10.1103/PhysRevA.85.062120",
    journal = "Phys. Rev. A",
    volume = "85",
    pages = "062120",
    year = "2012"
}

@article{Griffiths:1984rx,
    author = "Griffiths, Robert B.",
    title = "{Consistent histories and the interpretation of quantum mechanics}",
    doi = "10.1007/BF01015734",
    journal = "J. Statist. Phys.",
    volume = "36",
    pages = "219--272",
    year = "1984"
}

@article{Omnes:1992ag,
    author = "Omnes, Roland",
    title = "{Consistent interpretations of quantum mechanics}",
    doi = "10.1103/RevModPhys.64.339",
    journal = "Rev. Mod. Phys.",
    volume = "64",
    pages = "339--382",
    year = "1992"
}

@article{Gell-Mann:1992wkv,
    author = "Gell-Mann, Murray and Hartle, James B.",
    title = "{Classical equations for quantum systems}",
    eprint = "gr-qc/9210010",
    archivePrefix = "arXiv",
    reportNumber = "UCSBTH-91-15",
    doi = "10.1103/PhysRevD.47.3345",
    journal = "Phys. Rev. D",
    volume = "47",
    pages = "3345--3382",
    year = "1993"
}

@article{carroll2021quantum-singh,
  title={Quantum mereology: Factorizing {H}ilbert space into subsystems with quasiclassical dynamics},
  author={Carroll, Sean M and Singh, Ashmeet},
  journal={Physical Review A},
  volume={103},
  number={2},
  pages={022213},
  year={2021},
  publisher={APS}
}

@article{deutsch1985quantum-interpretation-basis,
  title={Quantum theory as a universal physical theory},
  author={Deutsch, David},
  journal={International Journal of Theoretical Physics},
  volume={24},
  pages={1--41},
  year={1985},
  publisher={Springer}
}

@article{Anglin:1996bb,
    author = "Anglin, J. R. and Paz, J. P. and Zurek, W. H.",
    title = "{Deconstructing decoherence}",
    eprint = "quant-ph/9611045",
    archivePrefix = "arXiv",
    reportNumber = "LA-UR-96-2230, LA-UR 96-2230",
    doi = "10.1103/PhysRevA.55.4041",
    journal = "Phys. Rev. A",
    volume = "55",
    pages = "4041",
    year = "1997"
}

@misc{Stoica:2024nhh,
      title={Does the {H}amiltonian determine the tensor product structure and the 3d space?}, 
      author={Ovidiu Cristinel Stoica},
      year={2024},
      eprint={2401.01793},
      archivePrefix={arXiv},
      primaryClass={quant-ph}
}

@article{Cotler:2017abq,
    author = "Cotler, Jordan S. and Penington, Geoffrey R. and Ranard, Daniel H.",
    title = "{Locality from the Spectrum}",
    eprint = "1702.06142",
    archivePrefix = "arXiv",
    primaryClass = "quant-ph",
    reportNumber = "SU-ITP-17-01",
    doi = "10.1007/s00220-019-03376-w",
    journal = "Commun. Math. Phys.",
    volume = "368",
    number = "3",
    pages = "1267--1296",
    year = "2019"
}

@article{Albrecht:1992rs,
    author = "Albrecht, Andreas",
    title = "{Investigating decoherence in a simple system}",
    reportNumber = "FERMILAB-PUB-91-101-A-REV, FERMILAB-PUB-91-101-A",
    doi = "10.1103/PhysRevD.46.5504",
    journal = "Phys. Rev. D",
    volume = "46",
    pages = "5504--5520",
    year = "1992"
}

@inbook{Lidar_2003,
   title={Decoherence-Free Subspaces and Subsystems},
   ISBN={9783540448747},
   ISSN={0075-8450},
   url={http://dx.doi.org/10.1007/3-540-44874-8_5},
   DOI={10.1007/3-540-44874-8_5},
   booktitle={Lecture Notes in Physics},
   publisher={Springer Berlin Heidelberg},
   author={Lidar, Daniel A. and Birgitta Whaley, K.},
   year={2003},
   pages={83–120} }

@article{Dowker:1994ac,
    author = "Dowker, Fay and Kent, Adrian",
    title = "{Properties of consistent histories}",
    eprint = "gr-qc/9409037",
    archivePrefix = "arXiv",
    reportNumber = "DAMTP-94-66A",
    doi = "10.1103/PhysRevLett.75.3038",
    journal = "Phys. Rev. Lett.",
    volume = "75",
    pages = "3038--3041",
    year = "1995"
}

@article{Albrecht:2021sic,
    author = "Albrecht, Andreas and Baunach, Rose and Arrasmith, Andrew",
    title = "{Einselection, equilibrium, and cosmology}",
    eprint = "2105.14017",
    archivePrefix = "arXiv",
    primaryClass = "hep-th",
    doi = "10.1103/PhysRevD.106.123507",
    journal = "Phys. Rev. D",
    volume = "106",
    number = "12",
    year = "2022"
}

@article{Stoica:2021rqi,
    author = "Stoica, Ovidiu Cristinel",
    title = "{3D-space and the preferred basis cannot uniquely emerge from the quantum structure}",
    eprint = "2102.08620",
    archivePrefix = "arXiv",
    primaryClass = "quant-ph",
    doi = "10.4310/ATMP.2022.v26.n10.a12",
    journal = "Adv. Theor. Math. Phys.",
    volume = "26",
    number = "10",
    pages = "3895--3962",
    year = "2022"
}

@misc{Stoica:2021bub,
      title={Refutation of {H}ilbert Space Fundamentalism}, 
      author={Ovidiu Cristinel Stoica},
      year={2022},
      eprint={2103.15104},
      archivePrefix={arXiv},
      primaryClass={quant-ph}
}

@article{zanardi2022operational,
  title={Operational Quantum Mereology and Minimal Scrambling},
  author={Zanardi, Paolo and Dallas, Emanuel and Lloyd, Seth},
  journal={arXiv preprint arXiv:2212.14340},
  year={2022}
}

@misc{knutson2000honeycombs,
      title={Honeycombs and sums of {H}ermitian matrices}, 
      author={Allen Knutson and Terence Tao},
      year={2000},
      eprint={math/0009048},
      archivePrefix={arXiv},
      primaryClass={math.RT}
}

@article{zanardi2004quantum-lloyd-lidar,
  title={Quantum tensor product structures are observable induced},
  author={Zanardi, Paolo and Lidar, Daniel A and Lloyd, Seth},
  journal={Physical review letters},
  volume={92},
  number={6},
  pages={060402},
  year={2004},
  publisher={APS}
}

@article{zanardi2001virtual-subsystems,
  title={Virtual quantum subsystems},
  author={Zanardi, Paolo},
  journal={Physical Review Letters},
  volume={87},
  number={7},
  pages={077901},
  year={2001},
  publisher={APS}
}

@article{arrasmith2019variational,
  title={Variational consistent histories as a hybrid algorithm for quantum foundations},
  author={Arrasmith, Andrew and Cincio, Lukasz and Sornborger, Andrew T and Zurek, Wojciech H and Coles, Patrick J},
  journal={Nature communications},
  volume={10},
  number={1},
  pages={3438},
  year={2019},
  publisher={Nature Publishing Group UK London}
}

@article{piazza2010glimmers,
  title={Glimmers of a pre-geometric perspective},
  author={Piazza, Federico},
  journal={Foundations of Physics},
  volume={40},
  pages={239--266},
  year={2010},
  publisher={Springer}
}

@article{tegmark2015consciousness,
  title={Consciousness as a state of matter},
  author={Tegmark, Max},
  journal={Chaos, Solitons \& Fractals},
  volume={76},
  pages={238--270},
  year={2015},
  publisher={Elsevier}
}

@book{schlosshauer2007decoherence,
  title={Decoherence and the Quantum-To-Classical Transition},
  author={Schlosshauer, M.A.},
  isbn={9783540357735},
  lccn={2007930038},
  series={The Frontiers Collection},
  url={https://books.google.co.uk/books?id=1qrJUS5zNbEC},
  year={2007},
  publisher={Springer}
}

@article{Albrecht:1992uc,
    author = "Albrecht, Andreas",
    title = "{Following a `collapsing' wave function}",
    eprint = "hep-th/9309051",
    archivePrefix = "arXiv",
    reportNumber = "FERMILAB-PUB-92-318-A, IMPERIAL-TP-92-93-03",
    doi = "10.1103/PhysRevD.48.3768",
    journal = "Phys. Rev. D",
    volume = "48",
    pages = "3768--3778",
    year = "1993"
}

@article{Albrecht:2007mm,
    author = "Albrecht, Andreas and Iglesias, Alberto",
    title = "{The Clock ambiguity and the emergence of physical laws}",
    eprint = "0708.2743",
    archivePrefix = "arXiv",
    primaryClass = "hep-th",
    doi = "10.1103/PhysRevD.77.063506",
    journal = "Phys. Rev. D",
    volume = "77",
    pages = "063506",
    year = "2008"
}

@article{schlosshauer2019quantum-dec-review,
  title={Quantum decoherence},
  author={Schlosshauer, Maximilian},
  journal={Physics Reports},
  volume={831},
  pages={1--57},
  year={2019},
  publisher={Elsevier}
}

@article{zurek1998decoherence-existential-interpretation-rough-guide,
  title={Decoherence, einselection and the existential interpretation (the rough guide)},
  author={Zurek, Wojciech H},
  journal={Philosophical Transactions of the Royal Society of London. Series A: Mathematical, Physical and Engineering Sciences},
  volume={356},
  number={1743},
  pages={1793--1821},
  year={1998},
  publisher={The Royal Society}
}

@article{strasberg2023everything,
  title={Everything Everywhere All At Once: A First Principles Numerical Demonstration of Emergent Decoherent Histories},
  author={Strasberg, Philipp and Reinhard, Teresa E and Schindler, Joseph},
  journal={arXiv preprint arXiv:2304.10258},
  year={2023}
}

@article{Adil2023entanglement,
   title={Entanglement masquerading in the {CMB}},
   volume={2023},
   ISSN={1475-7516},
   url={http://dx.doi.org/10.1088/1475-7516/2023/06/024},
   DOI={10.1088/1475-7516/2023/06/024},
   number={06},
   journal={Journal of Cosmology and Astroparticle Physics},
   publisher={IOP Publishing},
   author={Adil, Arsalan and Albrecht, Andreas and Baunach, Rose and Holman, R. and Ribeiro, Raquel H. and Richard, Benoit J.},
   year={2023},
   month=jun, pages={024} }

@article{zurek1993coherent-states-Habib-Paz,
  title={Coherent states via decoherence},
  author={Zurek, Wojciech H and Habib, Salman and Paz, Juan Pablo},
  journal={Physical Review Letters},
  volume={70},
  number={9},
  pages={1187},
  year={1993},
  publisher={APS}
}

@article{albrecht2023adaptedcalmodel,
  title={Adapted {C}aldeira-{L}eggett Model},
  author={Albrecht, Andreas and Baunach, Rose and Arrasmith, Andrew},
  journal={Physical Review Research},
  volume={5},
  number={2},
  pages={023187},
  year={2023},
  publisher={APS}
}

@article{paz1993reduction-zurek-habib,
  title={Reduction of the wave packet: Preferred observable and decoherence time scale},
  author={Paz, Juan Pablo and Habib, Salman and Zurek, Wojciech H},
  journal={Physical Review D},
  volume={47},
  number={2},
  pages={488},
  year={1993},
  publisher={APS}
}

@book{zurek2025decoherence,
  title     = {Decoherence and Quantum Darwinism: From Quantum Foundations to Classical Reality},
  author    = {Wojciech Hubert Zurek},
  year      = {2025},
  publisher = {Cambridge University Press},
  address   = {Cambridge},
  isbn      = {9781009552868},
  doi       = {10.1017/9781009552868}
}

@article{zurek1981pointer,
  title={Pointer basis of quantum apparatus: Into what mixture does the wave packet collapse?},
  author={Zurek, Wojciech H},
  journal={Physical Review D},
  volume={24},
  number={6},
  pages={1516},
  year={1981},
  publisher={APS}
}

@article{joos1985emergence-zeh,
  title={The emergence of classical properties through interaction with the environment},
  author={Joos, Eric and Zeh, H Dieter},
  journal={Zeitschrift f{\"u}r Physik B Condensed Matter},
  volume={59},
  pages={223--243},
  year={1985},
  publisher={Springer}
}

@article{Paz_1999,
   title={Quantum Limit of Decoherence: Environment Induced Superselection of Energy Eigenstates},
   volume={82},
   ISSN={1079-7114},
   url={http://dx.doi.org/10.1103/PhysRevLett.82.5181},
   DOI={10.1103/physrevlett.82.5181},
   number={26},
   journal={Physical Review Letters},
   publisher={American Physical Society (APS)},
   author={Paz, Juan Pablo and Zurek, Wojciech Hubert},
   year={1999},
   month=jun, pages={5181–5185} }

@article{ollivier2004objective,
  title={Objective properties from subjective quantum states: Environment as a witness},
  author={Ollivier, Harold and Poulin, David and Zurek, Wojciech H},
  journal={Physical Review Letters},
  volume={93},
  number={22},
  pages={220401},
  year={2004},
  publisher={APS}
}

@article{ollivier2005environment,
  title={Environment as a witness: Selective proliferation of information and emergence of objectivity in a quantum universe},
  author={Ollivier, Harold and Poulin, David and Zurek, Wojciech H},
  journal={Physical Review A},
  volume={72},
  number={4},
  pages={042113},
  year={2005},
  publisher={APS}
}

@article{zurek2009quantum-darwinism,
  title={Quantum darwinism},
  author={Zurek, Wojciech Hubert},
  journal={Nature Physics},
  volume={5},
  number={3},
  pages={181--188},
  year={2009},
  publisher={Nature Publishing Group UK London}
}

@article{zurek2003decoherence-and-the-quantum-origins,
  title={Decoherence, einselection, and the quantum origins of the classical},
  author={Zurek, Wojciech Hubert},
  journal={Reviews of modern physics},
  volume={75},
  number={3},
  pages={715},
  year={2003},
  publisher={APS}
}

@article{martineau2007decoherence,
  title={On the decoherence of primordial fluctuations during inflation},
  author={Martineau, Patrick},
  journal={Classical and Quantum Gravity},
  volume={24},
  number={23},
  pages={5817},
  year={2007},
  publisher={IOP Publishing}
}

@article{burgess2006decoherence-holman,
  title={On the decoherence of primordial fluctuations during inflation},
  author={Burgess, CP and Holman, R and Hoover, D},
  journal={arXiv preprint astro-ph/0601646},
  year={2006}
}

@article{burgess2023minimal-holman,
  title={Minimal decoherence from inflation},
  author={Burgess, CP and Holman, R and Kaplanek, Greg and Martin, Jerome and Vennin, Vincent},
  journal={Journal of Cosmology and Astroparticle Physics},
  volume={2023},
  number={07},
  pages={022},
  year={2023},
  publisher={IOP Publishing}
}

@article{kiefer1998emergence,
  title={Emergence of classicality for primordial fluctuations: Concepts and analogies},
  author={Kiefer, Claus and Polarski, David},
  journal={Annalen der Physik},
  volume={510},
  number={3},
  pages={137--158},
  year={1998},
  publisher={Wiley Online Library}
}

@article{sharman2007decoherence,
  title={Decoherence due to the Horizon after Inflation},
  author={Sharman, Jonathan W and Moore, Guy D},
  journal={Journal of Cosmology and Astroparticle Physics},
  volume={2007},
  number={11},
  pages={020},
  year={2007},
  publisher={IOP Publishing}
}

@article{polarski1996semiclassicality-starobinsky-decoherence-wo-decoherence,
  title={Semiclassicality and decoherence of cosmological perturbations},
  author={Polarski, David and Starobinsky, Alexei A},
  journal={Classical and Quantum Gravity},
  volume={13},
  number={3},
  pages={377},
  year={1996},
  publisher={IOP Publishing}
}

@article{albrecht1994inflation-squeezing,
  title={Inflation and squeezed quantum states},
  author={Albrecht, Andreas and Ferreira, Pedro and Joyce, Michael and Prokopec, Tomislav},
  journal={Physical Review D},
  volume={50},
  number={8},
  pages={4807},
  year={1994},
  publisher={APS}
}

@article{everett1957relative,
  title={``{R}elative state'' formulation of quantum mechanics},
  author={Everett III, Hugh},
  journal={Reviews of Modern Physics},
  volume={29},
  number={3},
  pages={454},
  year={1957},
  publisher={APS}
}

@phdthesis{everett1956Thesis,
    author = "Everett, III, Hugh",
    title = "{The Theory of the Universal Wave Function}",
    school = "Princeton U.",
    year = "1956"
}

@article{wheeler1957assessment,
  title={Assessment of {E}verett's" relative state" formulation of quantum theory},
  author={Wheeler, John A},
  journal={Reviews of Modern Physics},
  volume={29},
  number={3},
  pages={463},
  year={1957},
  publisher={APS}
}

@article{dewitt1973many-worlds,
  title={The many-universes interpretation of quantum mechanics},
  author={DeWitt, Bryce S},
  journal={The many-worlds interpretation of quantum mechanics (Journal)},
  pages={167},
  year={1973}
}

@article{dewitt1970quantum-Physics-Today,
  title={Quantum mechanics and reality},
  author={DeWitt, Bryce S},
  journal={Physics Today},
  volume={23},
  number={9},
  pages={30--35},
  year={1970}
}

@book{wallace2012emergent-multiverse,
  title={The emergent multiverse: Quantum theory according to the {E}verett interpretation},
  author={Wallace, David},
  year={2012},
  publisher={Oxford University Press, USA},
}

@article{heunisch2023tunable-hartmann,
  title={Tunable coupler to fully decouple superconducting qubits},
  author={Heunisch, Lukas and Eichler, Christopher and Hartmann, Michael J},
  journal={arXiv preprint arXiv:2306.17007},
  year={2023}
}

@article{quiroz2024dynamically-lidar-DFS,
doi = {10.1088/1361-6633/ad6805},
url = {https://dx.doi.org/10.1088/1361-6633/ad6805},
year = {2024},
month = {aug},
publisher = {IOP Publishing},
volume = {87},
number = {9},
pages = {097601},
author = {Quiroz, Gregory and Pokharel, Bibek and Boen, Joseph and Tewala, Lina and Tripathi, Vinay and Williams, Devon and Wu, Lian-Ao and Titum, Paraj and Schultz, Kevin and Lidar, Daniel},
title = {Dynamically generated decoherence-free subspaces and subsystems on superconducting qubits},
journal = {Reports on Progress in Physics}
}

@article{higham1989matrix,
  title={Matrix nearness problems and applications},
  author={Higham, Nicholas J},
  journal={Applications of matrix theory},
  volume={22},
  year={1989}
}

@article{green1952orthogonal,
  title={The orthogonal approximation of an oblique structure in factor analysis},
  author={Green, Bert F},
  journal={Psychometrika},
  volume={17},
  number={4},
  pages={429--440},
  year={1952},
  publisher={Springer}
}

@article{schollwock2005density,
  title={The density-matrix renormalization group},
  author={Schollw{\"o}ck, Ulrich},
  journal={Reviews of Modern Physics},
  volume={77},
  number={1},
  pages={259},
  year={2005},
  publisher={APS}
}

@article{foster1988recent-interpretation-basis,
  title={On a recent attempt to define the interpretation basis in the many worlds interpretation of quantum mechanics},
  author={Foster, Sara and Brown, Harvey},
  journal={International Journal of Theoretical Physics},
  volume={27},
  pages={1507--1531},
  year={1988},
  publisher={Springer}
}

@article{mansuroglu2024quantum,
  title={Quantum Tensor Product Decomposition from {C}hoi State Tomography},
  author={Mansuroglu, Refik and Adil, Arsalan and Hartmann, Michael J and Holmes, Zo{\"e} and Sornborger, Andrew T},
  journal={arXiv preprint arXiv:2402.05018},
  year={2024}
}

@article{coleman1975quantum-bosonization,
  title={Quantum sine-{G}ordon equation as the massive {T}hirring model},
  author={Coleman, Sidney},
  journal={Physical Review D},
  volume={11},
  number={8},
  pages={2088},
  year={1975},
  publisher={APS}
}

@article{lieb1961two-XY_JW,
  title={Two soluble models of an antiferromagnetic chain},
  author={Lieb, Elliott and Schultz, Theodore and Mattis, Daniel},
  journal={Annals of Physics},
  volume={16},
  number={3},
  pages={407--466},
  year={1961},
  publisher={Elsevier}
}

@article{schultz1964two-lieb-Ising_JW,
  title={Two-dimensional {I}sing model as a soluble problem of many fermions},
  author={Schultz, Theodore D and Mattis, Daniel C and Lieb, Elliott H},
  journal={Reviews of Modern Physics},
  volume={36},
  number={3},
  pages={856},
  year={1964},
  publisher={APS}
}

@article{von1998bosonization-reformionization,
  title={Bosonization for beginners—refermionization for experts},
  author={Von Delft, Jan and Schoeller, Herbert},
  journal={Annalen der Physik},
  volume={510},
  number={4},
  pages={225--305},
  year={1998},
  publisher={Wiley Online Library}
}

@article{wallace2002worlds-consistent-histories,
  title={Worlds in the Everett interpretation},
  author={Wallace, David},
  journal={Studies in History and Philosophy of Science Part B: Studies in History and Philosophy of Modern Physics},
  volume={33},
  number={4},
  pages={637--661},
  year={2002},
  publisher={Elsevier}
}

@article{pokharel2018demonstration-lidar,
  title={Demonstration of fidelity improvement using dynamical decoupling with superconducting qubits},
  author={Pokharel, Bibek and Anand, Namit and Fortman, Benjamin and Lidar, Daniel A},
  journal={Physical review letters},
  volume={121},
  number={22},
  pages={220502},
  year={2018},
  publisher={APS}
}

@article{eakins2002factorization-microsingularity,
  title={Factorization and entanglement in quantum systems},
  author={Eakins, Jon and Jaroszkiewicz, George},
  journal={Journal of Physics A: Mathematical and General},
  volume={36},
  number={2},
  pages={517},
  year={2002},
  publisher={IOP Publishing}
}

@article{bunch1978quantum,
  title={Quantum field theory in de {S}itter space: renormalization by point-splitting},
  author={Bunch, Timothy S and Davies, Paul CW},
  journal={Proceedings of the Royal Society of London. A. Mathematical and Physical Sciences},
  volume={360},
  number={1700},
  pages={117--134},
  year={1978},
  publisher={The Royal Society London}
}

\appendix

\medskip

\section{ \edits{Definitions}}
\label{sec:appendix-Definitions}
In this paper we use some words that are utilized differently by different sectors of our community. 
Here we clarify the usages we chose for this paper.

\begin{description}
     \item[Subsystem] If a Hilbert space has a tensor product structure, e.g.
     \begin{equation}
      \mathcal{H}=\mathcal{A}\otimes\mathcal{B}        
     \end{equation}
     the factors $\mathcal{A}$ and $\mathcal{B}$ are candidates to be considered \emph{subsystems} of the full quantum system. In practice it is helpful to consider additional properties (such as predictable behavior) that extend beyond this formal construction. That is the focus of this paper. Equivalent statements can be made for tensor product structures formed of any number of factors.  We note that such constructions require that the dimension of $\mathcal{H}$ equals the product of the dimensions of its factors, which means not all Hilbert spaces can be factored in all possible ways.

     \item[Quantum coherence] 
     Any system in a pure quantum state  can be thought of as having quantum coherence.  In practice quantum coherence gets more attention in some situations than others.  For example a billiard ball in a pure wavepacket state (i.e. the Universe is in a product state of $\ket{ball}\otimes \ket{rest} $), is rarely explicitly regarded as quantum coherent, although if it is measured in a different basis quantum coherence would be manifest (i.e. through interference).  The double slit diffraction pattern (which is {\em typically} measured in a completely different basis) is known for its quantum coherence.  Further discussion along these lines can be found in~\cite{albrecht1994inflation-squeezing,Albrecht_1994}. 
     
    \item[Decoherence] In the most general sense, decoherence is the loss of quantum coherence (as defined above) due to the entanglement with some other physical system. Many discussions focus on the case where the other system is a large ``environment'', although that is not a requirement for a system to decohere.  Some communities may use decoherence to refer to the onset of classicality or, more specifically, to refer to einselection (defined below)~\cite{zurek1993preferred-predictability-sieve}. In this paper, we use einselection to refer to a particular decohering phenomenon. 
    There is also a conceptual and technical tool called ``decohering histories'', which has only a loose connection to decoherence as defined here.  We use the term ``consistent histories'' for that tool. 
    
    \item[Einselection] or ``environment induced superselection'' refers to the case when interaction with the environment selects certain system states, called the ``pointer states'', that remain stable (possibly only approximately) against entanglement~\cite{zurek1981pointer, zurek2003decoherence-and-the-quantum-origins,zurek2025decoherence}.  

    To clarify the distinction we are making between general decoherence and decoherence that leads to einselection, we consider a quantum system consisting of two systems $A$ and $B$ which evolve according to a Hamiltonian $H$ which is a \emph{random} and time independent Hermitian matrix in the full $A\otimes B$ space. For concreteness, let the factorization that partitions the two systems also be random. If the systems start in an initial product state then evolution according to the random $H$ will certainly entangle $A$ with $B$, i.e.
       \begin{equation} 
        e^{-i H t } |\psi\rangle_A |\phi \rangle_{B} = \sum_i \alpha_i(t) |a_i(t)\rangle |b_i(t)\rangle
      \label{eqn:ABproduct}
    \end{equation}
  where the entangled state on the right is written in the Schmidt basis and there is more than one non-zero $\alpha_i(t)$. Consequently, the reduced system density matrix on system $A$ evolves as,
  \begin{equation}
      \rho_A(t)= \sum_k \alpha_k^2 (t) |a_k (t) \rangle \langle a_k(t)| \, ,
      \label{eq:rhos_dec}
  \end{equation}
  where we use the fact that $\langle b_i(t)|b_j(t)\rangle = \delta_{ij}$ by the definition of the Schmidt basis. 
The two subsystems have decohered each other (Eq.~\eqref{eq:rhos_dec} shows a mixed state). However, the  random nature of $H$ means there will be no special system states $|\psi \rangle_A$ one could choose in Eq.~\eqref{eqn:ABproduct} that would be robust to entanglement (as defined by Eq.~\eqref{eq:ContCostB}). Indeed, this claim is supported by our numerics in Fig.~\ref{fig:algorithm_summary}. Thus, despite there being decoherence, there are no pointer states, and so no einselection, for this example.  
    
    That this example is rather formal serves to highlight the fact that in the real world examples of decoherence and einselection are observed in parallel (hence the terms often being conflated). However, when conducting this sort of foundational study we have found the distinction between the two concepts to be helpful. And, indeed, it is einselection rather than decoherence that determines the emergence of subsystems. 
    
    \item[Pointer States] States that are (possibly only approximately) robust to entanglement with the environment. Wavepackets of everyday classical objects have this property.  This robustness also makes pointer states {\em predictable} since their evolution can be tracked without considering the effects of entanglement. 
    
    \item[Predictability Sieve]
    An operational tool for determining the pointer states of a system by explicitly checking which states are the most robust to entanglement with the environment (and therefore behave predictably). 
    
    \item[Branching] A term used to reference the different terms of an entangled wavefunction. In the relative state interpretation, einselection singles out the choice of basis with respect to which the different outcomes of each branch can be understood.  

    \item[Realm] A factorization of the global Hilbert space into a system and an environment factor such that the system behaves quasi-classically. For this work, we have taken classicality to be associated with the existence of pointer states (see definition of Subsystem above). A given Hamiltonian can admit multiple realms, with no simple correspondences between the subsystems of the different realms.  
    
    
\end{description}
\section{Block Diagonal Hamiltonians}
\label{sec:appendix-BlockD}

Here we describe how any Hamiltonian may be block-diagonalized to find a factorization that admits pointer states. This treatment pertains particularly to the category of pointer states discussed in item $b$ of subsection~\ref{sec:analysis}. 

\paragraph*{The block-diagonalization procedure.}\label{ap:construct}
Very simply we \textit{define} the factorization of the world Hilbert space such that the energy eigenstates are separable,

\begin{gather*}
|1\rangle_w = |1\rangle_s |1\rangle_e\\
|2\rangle_w = |1\rangle_s |2\rangle_e\\
|3\rangle_w = |2\rangle_s |1\rangle_e\\
\vdots\\
|E_k\rangle_w = |i\rangle_s |j\rangle_e
\end{gather*}
where $k$ is an index from 1 to $d_w$ and $i$ and $j$ run from 1 to $d_s$ and $d_e$ respectively. Thus, the prescription is a trivial relabelling of the energy eigenstates which allows one to write the Hamiltonian in the following form, 
\begin{equation}
    \begin{aligned}
    \label{eq:blockD_app}
    H = \sum_{k=1}^{d_w} E_k |E_k\rangle \langle E_k| 
    &= \sum_{i=1}^{d_s} |i\rangle_s \langle i| \otimes \sum_{j=1}^{d_e} |j\rangle_e \langle j| E_{ij} \\
    &=  \sum_{i=1}^{d_s} |i\rangle_s \langle i| \otimes  \tilde{H}_e^{(i)}
    \end{aligned}
\end{equation}
where we have defined $\tilde{H}_e^{(i)} = \sum_j |j\rangle_e \langle j| E_{ij} $. 

\medskip
\paragraph*{Recipe to find pointer states.} The argument above shows that any Hamiltonian admits pointer states. However, for a practitioner
it might not be clear how one could use that argument to find and prepare pointer states. 
The claim that one can always find pointer states from this perspective is the claim that it is always possible to find a factorization, parameterized by the unitary $B$, and basis of system states $\{ \ket{i}_s \}_{i=1}^{d_s}$ such that $ \ket{\phi(t)}_{w} = B e^{- i H t} \ket{i}_s \ket{\psi}_e$ remains unentangled for all times. One construction to do so is as follows:

\begin{enumerate}
    \item The first step is to diagonalize $H$. That is, find an arbitrary unitary $W$ and diagonal matrix $D$ such that $H = W D W^\dagger$.
    
    \item The second step is to decompose $D$ into two subsystems. Given a $D$ of the form 
    $D = \sum_i \lambda_i |i\rangle\langle i|_{w}$ we can introduce a bipartite splitting in the computation basis. \\
    
    For example, for two qubits we have,
    \begin{equation}
        \begin{aligned}
            &|0\rangle \rightarrow |00\rangle \\
            &|1\rangle \rightarrow |01\rangle \\
            &|2\rangle \rightarrow |10\rangle \\
            &|3\rangle \rightarrow |11\rangle \, .
        \end{aligned}
    \end{equation}
    Thus we can write $D$ as $D = |0\rangle \langle 0 |_s \otimes H_e^{(0)} + |1\rangle \langle 1 |_s \otimes H_e^{(1)}$ where $H_e^{(0)} := \lambda_0 |0\rangle \langle 0 |_e + \lambda_1 |1\rangle \langle 1 |_e$ and $H_e^{(1)} := \lambda_2 |0\rangle \langle 0 |_e + \lambda_3 |1\rangle \langle 1 |_e$.\\

    \item The states that remain robust to entanglement in the factorization $B = W^\dagger$, are given by $W \ket{i}_s \ket{\psi}_e$ for any environment state $\ket{\psi}_e$. 
    
\end{enumerate}
To see that this construction works note that
\begin{equation}
   \begin{aligned}
        B e^{- i H t} W \ket{i}_s \ket{\psi}_e &= W^\dagger W e^{-i D t} W^\dagger W \ket{i}_s \ket{\psi}_e \\
        &=  e^{-i D t} \ket{i}_s \ket{\psi}_s \\
        &= \ket{i}_s e^{-i  H_e^{(i)}  t} \ket{\psi}_e \\
        &= \ket{i}_s \ket{\psi^{(i)}(t)}_e
   \end{aligned} 
\end{equation}
where $\ket{\psi(t)}_e = e^{-i  H_e^{(i)}  t} \ket{\psi}_e$. This state clearly remains unentangled for all times.

\medskip

\paragraph*{Consequences of this construction.}

\begin{figure*}
    \centering
    \includegraphics[width=0.95\textwidth]{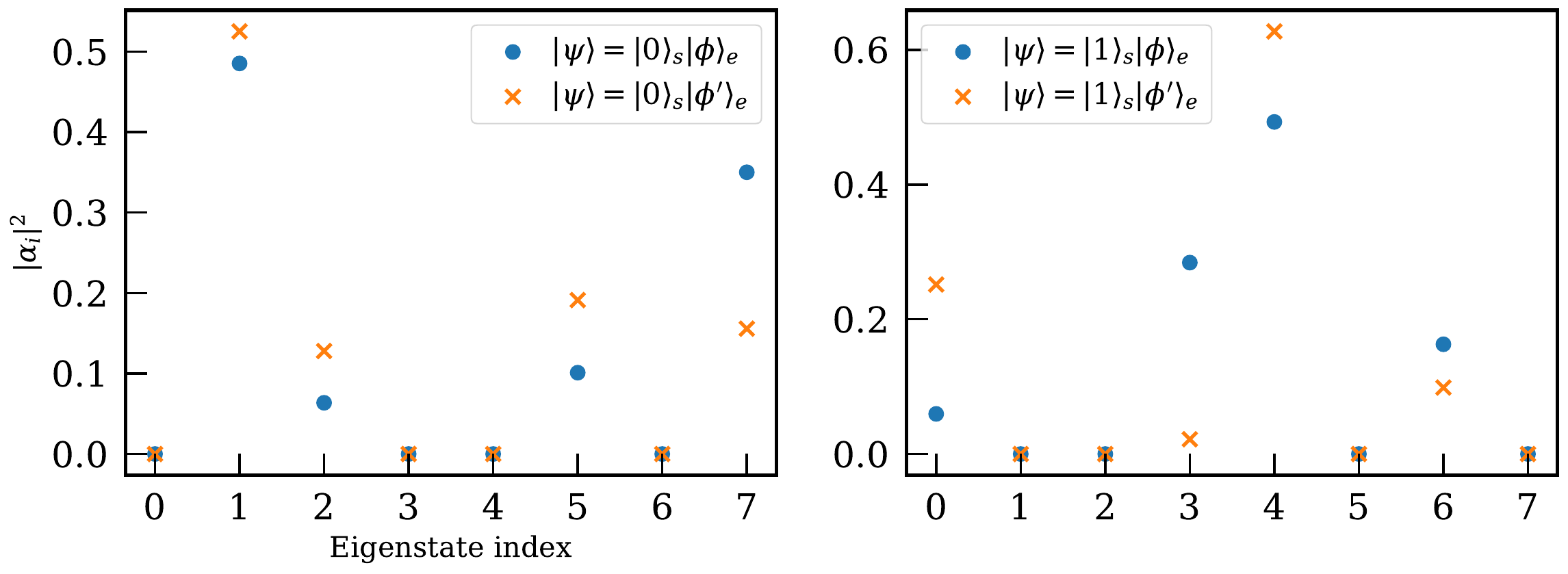}
    \caption{ Here we depict the effect of varying the environment state for a particular pointer state in a 1-qubit system and 2-qubit environment world. The figure shows the amplitudes of the coefficients of a product state composed of the pointer state and a random environment state in the energy eigenbasis. As discussed in the text, varying the environment state varies only the components in the subspace of eigenvectors that is projected by the particular choice of pointer state.}
    \label{fig:pointerH_eigcomps}
\end{figure*}

From this form of the Hamiltonian, it is easy to verify some observations. First, that the system states $|j\rangle_s$ are pointer states -- any initial product state where the system is in a pointer state remains a product state: $|j\rangle_s |\phi\rangle_e  \xrightarrow{U_t}  |j\rangle_s |\phi_j(t)\rangle_e$ where $ |\phi_j(t)\rangle_e = e^{-i \tilde{H}_e^{(j)} t } |\phi_j(t)\rangle_e $. Note that the system state is not dynamic and, though the environment state evolves in time, the evolution is confined within the subspace of eigenvectors that is projected by the particular choice of pointer state, as shown in Fig.~\ref{fig:pointerH_eigcomps}.

Finally, for an arbitrary pure initial state, the reduced system density matrix becomes diagonal in the pointer basis. To see this, start with an arbitrary product state $|\psi(0)\rangle_w = \sum_{j} \alpha_{j} |j\rangle_s |\phi\rangle_e$, 
(the assumption of an initial product state simplifies the calculation; the argument proceeds similarly for an arbitrary initial state). 
This evolves as 
\begin{equation}
    \begin{aligned}
        e^{-i H t} |\psi(0)\rangle_w = \sum_{j} \alpha_{j} |j\rangle_s  |\phi_j(t)\rangle_e
    \end{aligned}
\end{equation}
Thus the reduced state on the system is given by 
\begin{equation}
   \rho_s(t) = \sum_{jk} \alpha_{j}\alpha^*_k |j\rangle\langle k |_s c_{jk}(t)
\end{equation}
where $c_{jk}(t) =  \langle \phi_k(t)  |\phi_j(t)\rangle_e $. In the case that the global evolution drives the environment into approximately orthogonal states we have that $c_{jk}(t) \approx \delta_{jk}$ at large $t$ and so the system perfectly decoheres in the pointer basis, i.e.,
\begin{equation}
    \rho_s(t)  \rightarrow \sum_{j} |\alpha_{j}|^2 |j\rangle\langle j |_s \, .
\end{equation}

\section{Numerical Implementation}
\label{sec:numerical-implementation}
\subsection{Unitary optimization}
\label{subsec:appendix-numerical-imp-dmrg}
The optimization algorithm we employ is commonly used in the tensor network community. \edits{It is based on constructing the cost function $\Tr[\rho^2]$  (see Eq.~\eqref{eq:DisCostB}) as one tensor network contraction, where the \textit{network} here consists of several copies of the unitaries $A$ and $B$, $e^{-iHt}$ and $\psi\rangle_w$. By removing one of those tensors from the contraction, we compute its \textit{environment}, whose transpose represents the ideal local update for the removed tensor. This approach is, for example, the preferred method for optimizing local tensors in the \textit{density matrix renormalization group} (DMRG) algorithm~\cite{schollwock2005density}. We combine this with}
the so-called \textit{Singular Value Decomposition} (SVD) ``trick'', 
\edits{to find} the closest unitary matrix to a general matrix~\cite{green1952orthogonal}. 
\edits{The implementation of our algorithm with efficient matrix operations using the JAX library can be found on Github:~\href{github.com/MSRudolph/subsystems}{github.com/MSRudolph/subsystems}.}

To demonstrate the approach, assume a hypothetical cost function $\tilde{C} = 1- \langle \psi_1|U_1U_2|\psi_2\rangle$ with fixed states $\psi_1, \psi_2$ and trainable unitaries $U_1, U_2$. It is known that the optimal (non-unitary) operator $O^*_1$ in the position of $U_1$ relative to all other components being fixed can be calculated via $O^*_1 = |\psi_1\rangle\langle \psi_2|U^\dagger_2$. In other words, this is the operator that maximizes $\langle \psi_1|O_1U_2|\psi_2\rangle$ for any operator $O_1$. $O^*_1$ is commonly referred to as the \textit{environment tensor} of $U_1$ and is calculated by multiplying (in tensor network language called ``contracting'') all remaining terms after ``removing'' the term of interest (here $U_1$). The closest unitary $U^*_1$ to the non-unitary operator $O^*_1$ in terms of the Frobenius norm~\cite{higham1989matrix} can be calculated via the SVD, i.e., $O^*_1 = \mathcal{U}\mathcal{S}\mathcal{V}^\dagger$ and $U^*_1 = \mathcal{U}\mathcal{V}^\dagger$, where $\mathcal{U}, \mathcal{V}^\dagger$ are orthogonal matrices and $\mathcal{S}$ a diagonal matrix carrying the singular values. We can now do the same optimization step for $U_2$ and continuously iterate over both unitaries to minimize the cost $\tilde{C}$. Just as with any optimizer in non-convex landscapes, there is no guarantee that different initializations of $U_1$ and $U_2$ all converge to a global optimum where $\tilde{C} = 0$.

Now consider the cost function in Eq.~\eqref{eq:DisCostB}, which is not linear in the unitaries to be optimized, but is, in fact, quartic. In that case, the environment tensor that stems from removing one appearance of the unitaries denoted as $B$ is no longer the optimal operator in place of all appearances of $B$, but optimization can still be very successful in practice. The same approach can be taken for the unitary that prepares the initial input state (see Sec.~\ref{subsec:SR-initial-conds}). Optimization is usually improved by introducing a \textit{learning rate} that interpolates between the former and proposed unitaries. We apply decaying learning rates for the unitaries, which add fractions of the current unitaries to the environment tensor before the SVD. This results in a smaller step in the space of unitaries into the direction of the approximately optimal unitary per iteration. We note that the particular optimization protocol may be fine-tuned to find global optima of the cost function more often, but this is not the focus of our work.

\subsection{Training Times}
\label{subsec:appendix-numerical-imp-training-times}

\begin{figure*}
    \centering
    \includegraphics[width=0.49\linewidth]{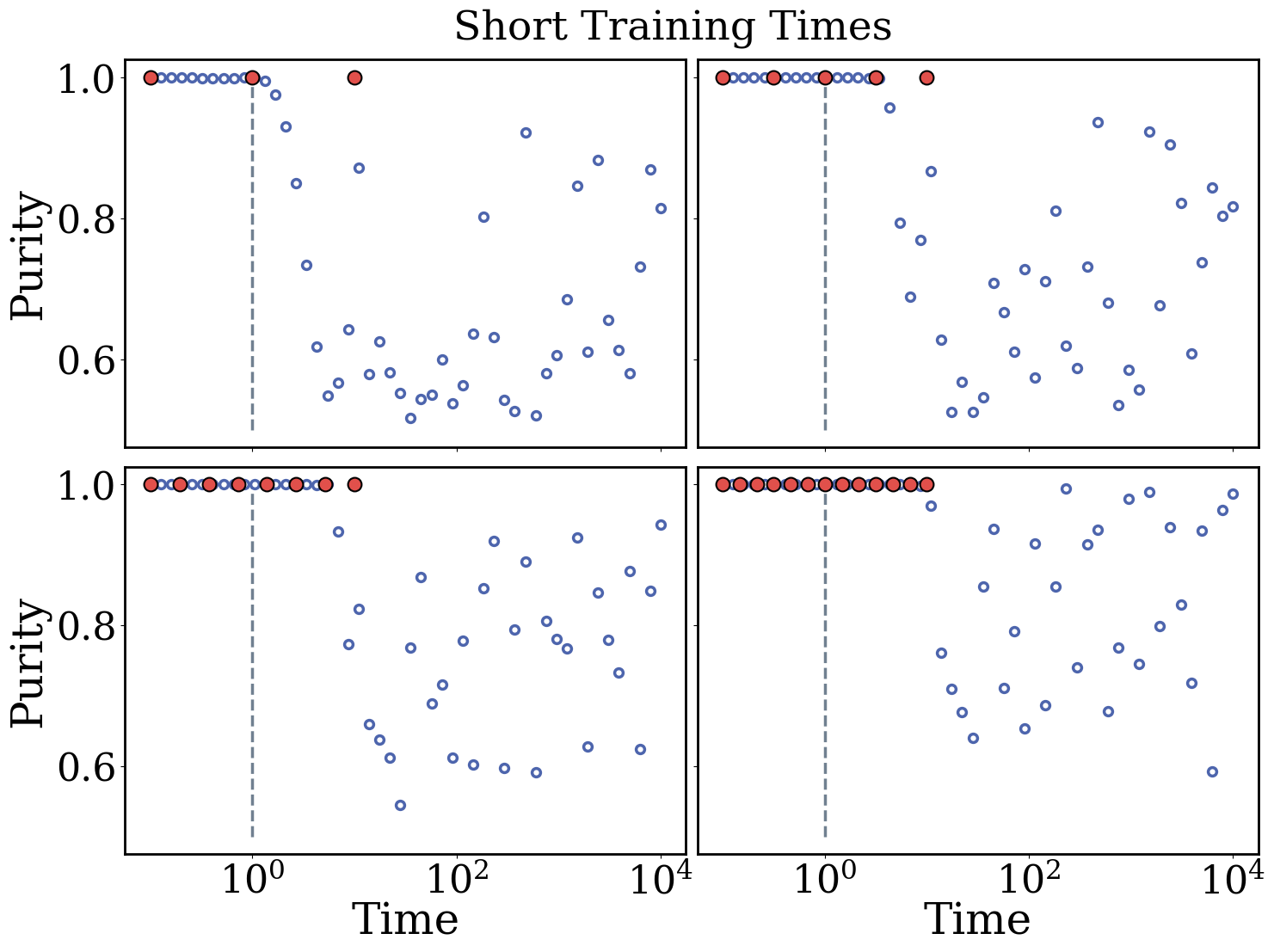}
    \includegraphics[width=0.49\linewidth]{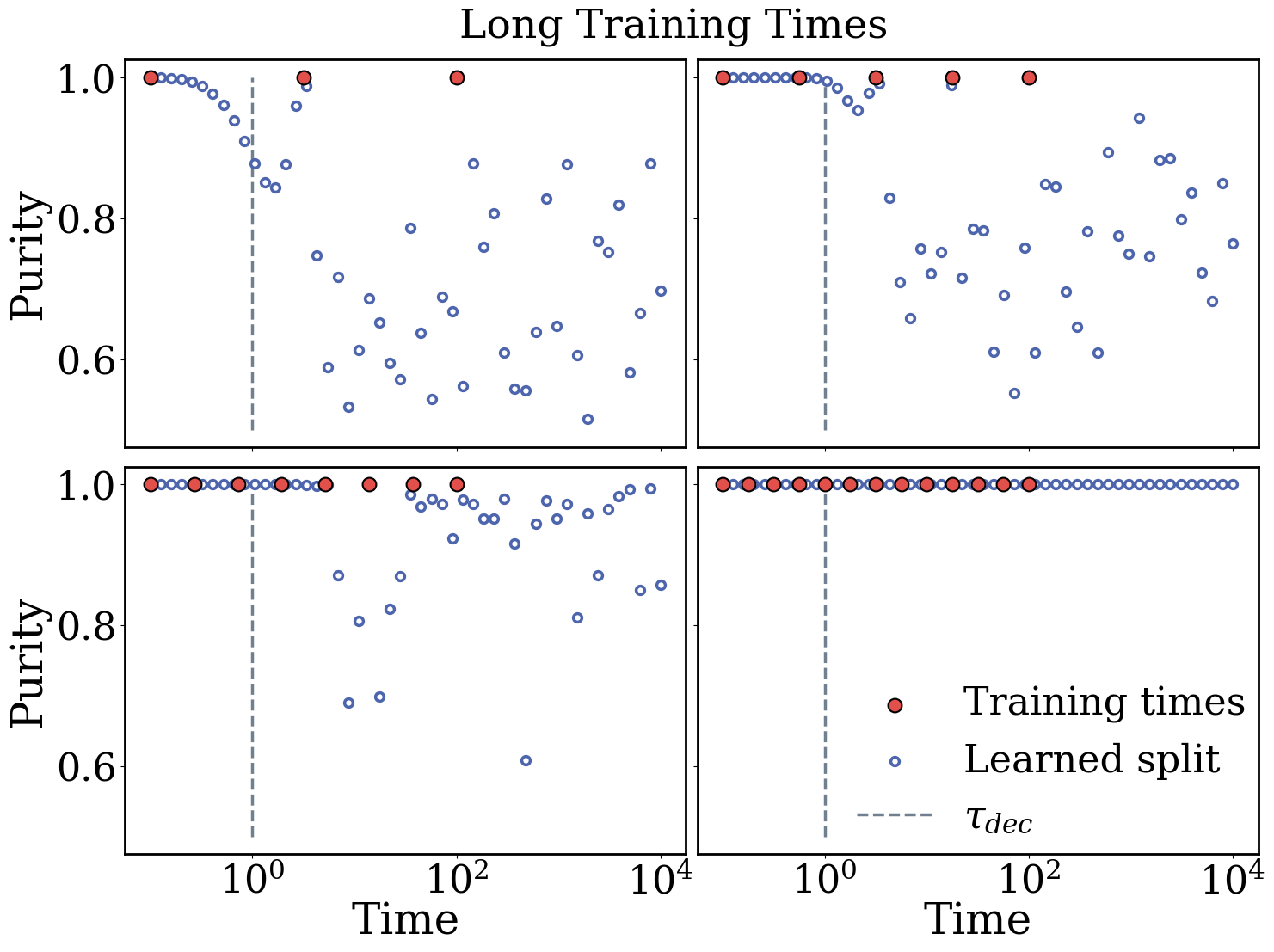}
    \caption{
    We illustrate the impact of different arrays of training times (solid red markers) on the full behavior of the solution found by our algorithm (open blue markers). For sets that extend sufficiently beyond the decoherence time (dashed line) and which are sufficiently dense, a well-behaved solution is found with unit purity (of the reduced density matrix) extending to late times (lower-rightmost panel).  In other cases solutions are found which fluctuate away from the desired pure evolution in a variety of ways (other panels).  The red training markers are separated by time $\Delta t$ and run to a maximum value $T_{\rm train}$ as per Eq.~\eqref{eq:DisCostB}.
    }
    \label{fig:training_times}
\end{figure*}

In order for the cost function Eq.~\eqref{eq:DisCostB} to be faithful, the number of time-steps should be large (empirically, we find this to be $k \gtrsim 2 d_w$ where $d_w \equiv \dim{(\mathcal{H}_w})$). This is because, for a small number of time steps, it is always possible to find specific states that minimize the cost at only the times on which the algorithm has been trained but that these states, unlike true pointer states, do not generalize to arbitrary times. This situation is explicitly depicted in the Fig.~\ref{fig:training_times} (left panel). One can visualize this cost landscape as one that has many local minima for small values of $k$, but that many of these ``evaporate'' as $k$ increases, leaving behind solutions that generalize to times outside the training set. 

Furthermore, for the faithfulness of the cost function, the set of training times needs to contain points sufficiently beyond the characteristic decoherence time of the Hamiltonian. This is important since the reduced density matrix of a generic initial product state will remain pure for $t\lesssim t_{\rm dec}$ where the decoherence time, $t_{\rm dec}$, is determined by $\lambda_{\rm max}$, the largest eigenvalue of the Hamiltonian, i.e. $t_{\rm dec} \equiv 1/ \lambda_{\rm max}$. This effect of varying $T_{\rm train}$ on the generalizability of the solution can also be seen in Fig.~\ref{fig:training_times} (right panel).

We note that it may be possible that if the ansatz for the unitary $B$ is chosen with care and in correspondence to the structure of the Hamiltonian, it is plausible that much fewer training times can be used. However, fully characterizing this algorithm is not the focus of this work.

\section{Detailed Statistics}
\label{sec:appendix-violin-plots}

\begin{figure*}
    \centering
    \begin{subfigure}{0.49\textwidth}
    \centering
    \includegraphics[width=\linewidth]{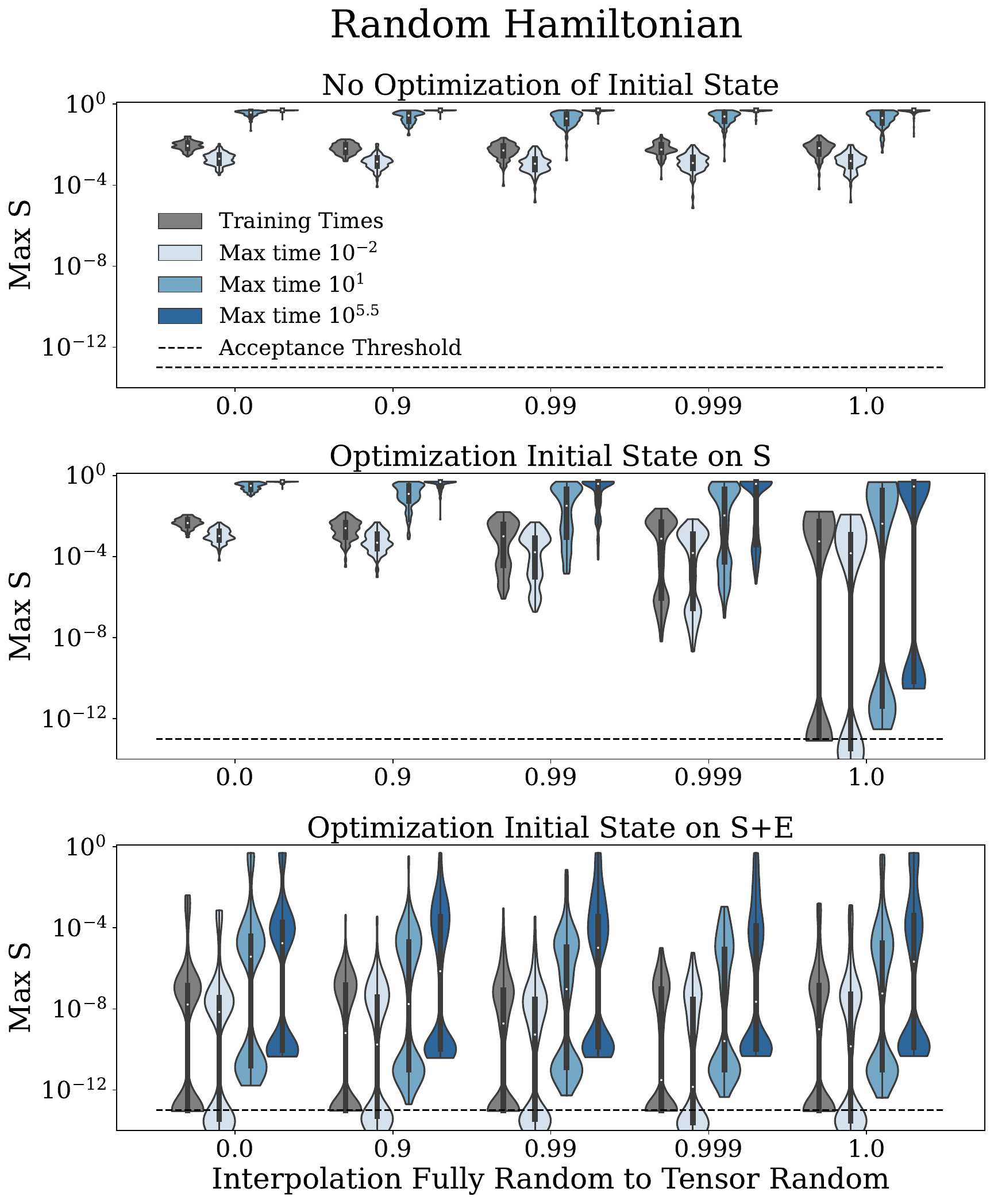}
    \end{subfigure}
    \hfill
    \begin{subfigure}{0.49\textwidth}
    \centering
    \includegraphics[width=\linewidth]{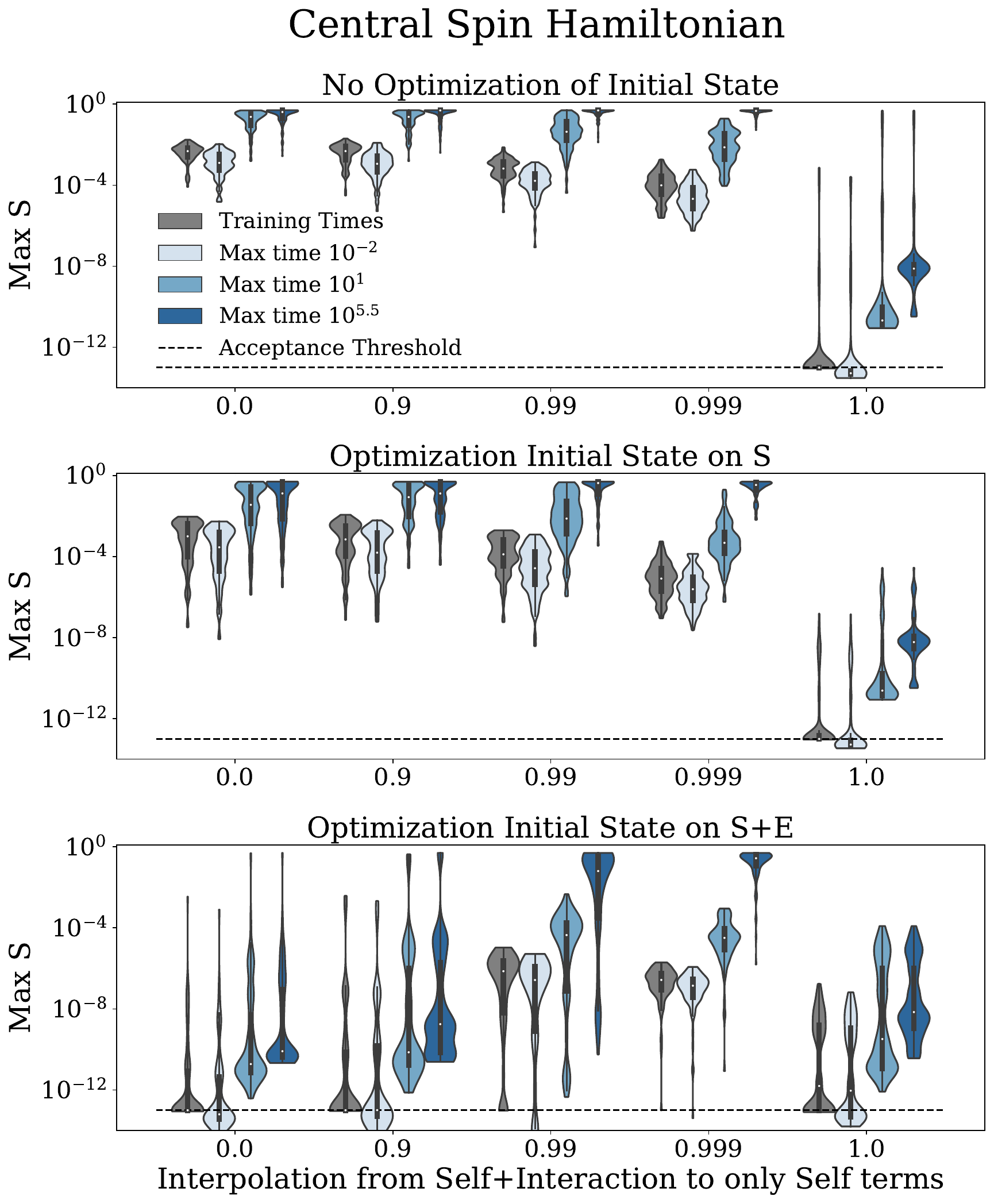}
    \end{subfigure}
    \caption{Violin plots capturing the statistics obtained by $100$ random initializations for each case. For each instance, the linear entropy of $\rho_s$ is evaluated for the set of training times (grey), as well as a set of times up to a maximum time (different shades of blue), and the \textit{highest} entropy (i.e. worst-case) point from this set is collected. The various violins then depict the details of the statistics summarized in Fig.~\ref{fig:algorithm_summary}. Their interpretation is discussed extensively in Appendix~\ref{sec:appendix-violin-plots}. }
    \label{fig:violin-plot}
\end{figure*}

\begin{figure}
    \centering    \includegraphics[width=\linewidth]{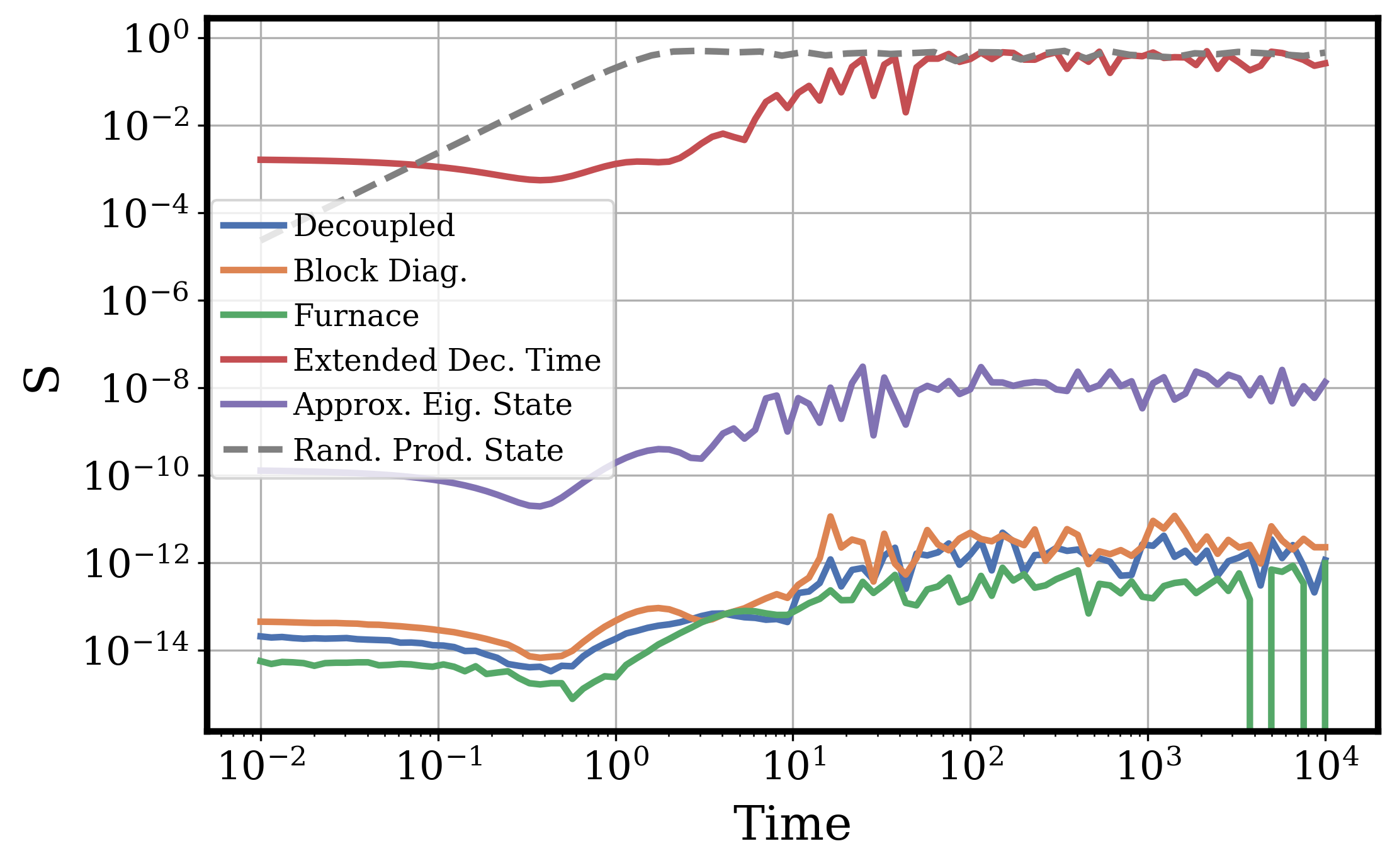}
    \caption{The figure depicts the linear entropy of the trained state (solid colors) for all of the classes of solutions discussed in Sec.~\ref{sec:analysis}. For comparison, we also show the evolution of the linear entropy of a random product state (dashed) in the fiducial factorization. All the solutions were obtained from the same fiducial Hamiltonian, the central spin (discussed in Sec.~\ref{subsec:SR-Hamiltonians-considered}), with one central and three environment qubits. This figure describes the ``quality'' of each of the solutions (i.e. the depth of the minimum on which the algorithm converges) and  should be compared with the last row of Fig.~\ref{fig:violin-plot}.   }
    \label{fig:lin_entropy_all_H}
\end{figure}

In this section, we provide the detailed numerical data behind Fig.~\ref{fig:algorithm_summary} in the main text. The data is depicted in Fig.~\ref{fig:violin-plot} in the form of \textit{violin plots}, where the width of the shapes indicates the density of points that have a certain entropy (defined in Eq.~\ref{eq:lin-entropy}). In the left panel of Fig.~\ref{fig:violin-plot}, we show the statistics obtained for the central spin model where the interaction term is slowly turned off (i.e. $\beta_i \rightarrow 0$ in Eq.~\eqref{eq:central-spin}). The right side depicts the random Hamiltonian case, where the interpolation goes from globally random Hamiltonians to a tensor product with a random Hamiltonian per qubit.

The purpose of this numerical analysis is to inspect the distribution of solutions, i.e., how often we achieve solutions with certain properties with our particular optimization algorithm. We optimize 100 randomly initialized instances and record the \textit{maximum} linear entropy of the reduced system density matrix until certain evolution times. We use two stopping criteria: either the instance achieves an ``acceptance threshold'' of $10^{-13}$ for the cost function Eq.~\eqref{eq:DisCostB}, or runs for $10^5$ iterations, whichever comes first. This is why instances for which the former stopping criterion is invoked cluster around the $10^{-13}$ level (see the plots labeled `Training Times' in e.g. the bottom row). These solutions also generalize well to late times. To see this, we show the maximum linear entropy until $t=10^{5.5}$. Note that we take $t=10^{5.5}$ as our ``late time'' limit, beyond which numerical instabilities can occur; if an instance performs well at this late time, we say that it ``generalizes''. In Fig.\ref{fig:lin_entropy_all_H}, we show that the decoupled, block diagonal, and furnace destination Hamiltonians discussed in Sec.~\ref{sec:analysis} can be identified as these lowest entropy solutions. 

In the third row of Fig.~\ref{fig:violin-plot}, there is a class of solutions for which the maximum training time entropy forms a ``blob'' away from the acceptance threshold at the $\mathcal{O}(10^{-7})$ level. Interestingly, these solutions continue to generalize well in the late time limit with linear entropy much less than one, though much greater than $\mathcal{O}(10^{-12})$. As shown in in Fig.~\ref{fig:lin_entropy_all_H},  the ``Approximate Eigenstate'' destination discussed in Sec.~\ref{sec:analysis} can be identified with this class of solution which appears to form a local minimum in the cost landscape. For the random Hamiltonian, it is also clear from the widths of these ``blobs'', relative to the width of the instances that cluster at the bottom, that these local minima solutions are just as numerous as those solutions which reach the acceptance threshold for the training times. However, in the \textit{weakly interacting} regime of the central spin (interpolation points $0.99, 0.999$), there are many more of these ``blob'' solutions with training cost $\approx \mathcal{O}(10^{-7})$. Given the description of the quality of each of the solutions in Fig.~\ref{fig:lin_entropy_all_H}, one might be tempted to conclude that all these mid-grade solutions fall in the ``Approximate Eigenstate'' category. However, because this is a numerical exploration, it is possible to have solutions that are close to one of the other destination Hamiltonians since the algorithm may trigger the maximum iteration stopping criterion before it can converge to the global minimum. Thus, some of these mid-grade solutions may be \textit{approximately} block diagonal or decoupled etc. We remind the reader again that this counting of the solutions is a reflection on the cost landscape and not (necessarily) a statement about a measure of \textit{physical} significance (see discussion in Sec.~\ref{sec:relative-state}). Thus, we interpret that the over-abundance of the ``blob'' solutions in the weakly interacting limit of the central spin occurs due to the cost landscape getting overwhelmed by many local minima which prevent the algorithm from converging to one of the (possible many) global minima.  

As discussed in Sec.~\ref{sec:analysis}, there is also the ``extended decoherence time'' class of solutions for which the cost is much less than one for the training times but that they do not generalize well in the late time limit. These appear as the `neck'' of the violins in the bottom row of Fig.~\ref{fig:violin-plot} and their slow growth in entropy (relative to a random product state) is depicted in Fig.~\ref{fig:lin_entropy_all_H}. Finally, a particularly intriguing case of the extended coherence time solutions can be seen in the middle row of Fig.~\ref{fig:violin-plot}. Here, the algorithm can achieve a low cost while optimizing the input state only on the \textit{fiducial} system qubit. Such solutions are rare, and it is unclear whether this finding has a physical interpretation or whether identifying such behavior with a quasi-classical subsystem perhaps akin to shapes found in the sky when cloud watching.

\end{document}